%% file: main.tex
\documentclass[runningheads]{llncs}

\usepackage{mathtools,calc}
\usepackage{tabularx}
\usepackage{latexsym}
\usepackage{amsfonts}
\usepackage{amssymb}
\usepackage{comment}
\usepackage[compatibility=false]{caption}
\usepackage{anyfontsize}
\usepackage{subcaption}
\usepackage{wrapfig}
\usepackage{paralist}
\usepackage{xspace}
\usepackage{graphicx}
\usepackage{booktabs}
\usepackage{multirow}
\usepackage{hhline}

\usepackage{titlecaps}
\usepackage{mathpartir}
\usepackage{microtype}

\usepackage[T1]{fontenc}
\usepackage[inference]{semantic}
\usepackage{mttex}

\usepackage{amsthm}
\usepackage{amstext}
\usepackage{thmtools}
\usepackage{stmaryrd}

\usepackage{listings}

\usepackage{macros}

\usepackage{hyperref}
\usepackage{placeins}

\def\@envspa{\hspace{0.3em}}
\def\@sa{\hspace{-0.2em}}
\def\@sb{\hspace{0.5em}}
\def\@sc{\hspace{-0.1em}}

\makeatletter
\lst@InstallKeywords k{attributes}{attributestyle}\slshape{attributestyle}{}ld
\makeatother

\definecolor{bluekeywords}{rgb}{0,0,1}
\definecolor{greencomments}{rgb}{0,0.5,0}
\definecolor{redstrings}{rgb}{0.64,0.08,0.08}
\definecolor{xmlcomments}{rgb}{0.5,0.5,0.5}
\definecolor{types}{rgb}{0.17,0.57,0.68}

\lstdefinelanguage{spyder}
{
  morekeywords={data,procedure,for, foreach, if, else, in, with, let, [], invariant, var},
  sensitive=false,
  morecomment=[l]{//},
  moreattributes={int, bool, []},
  attributestyle = \color{red}\bfseries
}

\lstdefinelanguage{boogie}
{
  morekeywords={function,global,procedure,while,if,else,requires,ensures,enforce,let,[],invariant, assume,assert, idx, len, min},
  sensitive=false,
  morecomment=[l]{//},
  moreattributes={int, bool, []},
  attributestyle = \color{red}\bfseries
}

\newcommand{\T}[1]{\mbox{\lstinline[language=spyder,columns=fixed,basicstyle=\lstmainstyle]^#1^}}

\lstMakeShortInline[columns=fixed]|

\title{Targeted Synthesis for Programming with Data Invariants}         

\author{
John Sarracino \and
Shraddha Barke \and
Hila Peleg \and
Sorin Lerner \and 
Nadia Polikarpova
}
\authorrunning{J. Sarracino et al.}

\institute{University of California, San Diego \\
\email{\{jsarraci,sbarke,hpeleg,npolikarpova,lerner\}@ucsd.edu}
}

\begin{document}

\maketitle
\input{abstract}

\input{intro}
\input{overview}
\input{language}
\input{synthesis}
\input{evaluation}

\input{related}

\bibliographystyle{splncs04}
{
\bibliography{library}
}

\input{appendix}

\end{document}

%% file: abstract.tex
\begin{abstract}
Data structures in a program are frequently subject to \emph{data invariants}---%
relations that must be maintained throughout program execution.
Traditionally, invariants are implicit, 
and are enforced by manually-crafted code.
Manual enforcement is error-prone, as the programmer must account 
for all locations that might break an invariant. 
Moreover, implicit invariants are brittle under code evolution: 
when the invariants and data structures change, 
the programmer must repeat the process of manually repairing
all of the code locations where invariants are violated.
In this work, we introduce \emph{programming with data invariants},
a new programming model, where invariants are exposed to the programmer as a language feature, 
and statically checked by the compiler.
Importantly, whenever programmer's code breaks an invariant, 
the compiler synthesizes a patch to restore it.
The two main challenges for implementing such a compiler
are to make patch synthesis efficient
and to avoid reverting changes made by the programmer.
To tackle these challenges, we introduce \technique,
an efficient patch synthesis algorithm,
which exploits structural similarity between invariants and code
to localize and simplify the synthesis problem. 
We evaluate our programming model and synthesis algorithm on a prototype language, \tool, 
which is a core imperative language with collections,
and supports a restricted but useful class of data invariants,
which we term \emph{iterator-based invariants}. 
We evaluate the succinctness and performance of \tool on a variety of programs inspired by web applications, 
and show that \tool allows concise specifications and implementation, 
and efficiently compiles and maintains data invariants.

\end{abstract}

%% file: intro.tex

\section{Introduction}
Programmers routinely face the task of enforcing \emph{data invariants}.
Prominent examples of data invariants include 
well-formedness of data structures,
model-view relations in interactive GUI applications,
and consistency between application data and the database.
Failure to properly enforce invariants is a common source of serious bugs and security vulnerabilities~\cite{BailisFFGHS15}.
Traditionally, programmers do not state invariants explicitly. 
Instead, they tacitly maintain invariants by sprinkling invariant-restoring snippets across their code. 
This ad-hoc practice 
is error-prone because the programmer must maintain a mental model
of which invariants are broken and how to restore them. 
In addition, these snippets are brittle under
software evolution: when data structures and their invariants change, 
the programmer must go over the entire code base to modify, remove, or add invariant-restoring snippets.


An attractive alternative to this traditional model 
is to let programmers state the desired invariants explicitly,
and have the programming language take responsibility for both \emph{checking} the invariant satisfaction, as well as \emph{enforcing} the invariants by updating the necessary data structures.
Static checking of invariants is the subject of much prior work in program verification~\cite{Barnett04,Dynamic,Friends,History,Muller06,Middelkoop08,Considerate,PolikarpovaTFM14,chargueraud2019gospel};
these techniques, however, can only identify the code locations where an invariant might be violated,
but they do not help the programmer restore the invariant.
On the other hand, \emph{declarative constraint programming}~\cite{rossi2006handbook,jaffar1992clp}
automatically adjusts program state to satisfy the invariant;
the downside, however, is that doing so at run time is both unpredictable and inefficient.
Wouldn't it be great if instead we could compile declarative constraints into imperative code?
Importantly, this would make the semantics of constraints more predictable,
since any ambiguity would have to be resolved at compile time,
when the compiler can ask the programmer for help.
In this work, we propose using \emph{program synthesis} technology
to compile declarative data invariants into imperative, invariant-enforcing patches.

Program synthesis is an active area of research~\cite{GulwaniHS12,Solar-Lezama:STTT13,TorlakB14,Polozov-Gulwani:OOPSLA15,Feng-al:PLDI17,Yaghmazadeh-al:OOPSLA17} 
that tackles the problem of generating programs from declarative constraints.
In particular, synthesis from logical specifications~\cite{leino2012program,Srivastava-al:POPL10,Kneuss-al:OOPSLA13,Delaware-al:POPL15,Polikarpova-al:PLDI16} 
takes as input a logical predicate over a program's inputs and outputs, 
and searches for a program that satisfies the predicate. 
We describe how program synthesis enables language support for data invariants through a motivating example.

\subsection{Motivating Example: Budget Planner}\label{sec:motivating-example}

\lstset{basicstyle=\lstfigstyle}
\begin{figure}[t]
  \centering
  \begin{subfigure}[b]{0.35\linewidth}
    \centering
    \includegraphics[width=0.8\columnwidth]{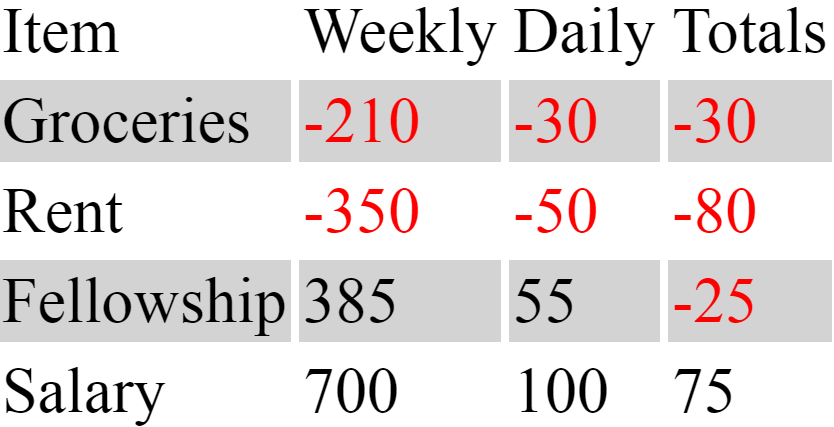}
    \caption{}
    \label{fig:budget-render}
  \end{subfigure}
  ~
  \begin{subfigure}[b]{0.6\linewidth}
  \centering
    \begin{lstlisting}%[basicstyle=\scriptsize]
var COLA = 1.05;
for (i = 0; i != rows.length; ++i) {
  var r = rows[i];
  if (r > 0) {
    r.day = r.day * COLA;
  }
  // Targeted synthesis:
  // assigns to 
  // r.week, r.total
}
// Naive synthesis: after loop
// regenerate rows and preserve days
\end{lstlisting}
    \caption{}
    \label{fig:budget-code}
  \end{subfigure}
  \caption{GUI application for building a budget from recurring expenses and incomes.}
    \label{fig:budget}
\end{figure}

Consider a budget builder application for recurring expenses and incomes, shown in \autoref{fig:budget}.
The amount for each item in the budget plan is stored in two different formats, \T{Weekly} and \T{Daily}, so that the end-user can provide input in the most relevant period.
For example, a budget for meals can be given in \T{Daily} units, rent can be given in \T{Weekly} units, etc.
To see whether the planned budget is balanced, the daily budget items are added up: 
a running total stored in \T{Totals}; the final entry of \T{Totals} contains
the expected overall surplus or deficit per day.
Revenues are distinguished from Expenses by rendering Revenues black and Expenses red.

Each of these application properties is a data invariant that the programmer 
has to maintain:
\begin{inparaenum}[(1)]
  \item \T{weekly} and \T{daily} are unit-conversions of each other,\label{unitconv-invariant}
  \item \T{totals} is a running sum of the \T{daily} values, and
  \item if an entry is negative, its font color is red.
\end{inparaenum}

Consider a function that adjusts the income in an existing budget according to a cost-of-living index.
This function, shown in \autoref{fig:budget-code}, multiplies each positive daily item by the Cost-of-Living-Adjustment (COLA) constant.
The loop in \autoref{fig:budget-code} breaks invariants (1) and (2): 
the \T{weekly} and \T{total} values are stale.
Our goal is to synthesize an \emph{invariant patch},
\ie a code snippet that, when inserted into the function body, will provably restore the broken data invariants.

At a first glance, it seems natural to insert the patch at the end of the function,
using the programmer-provided data invariants as the specification for synthesis.
Unfortunately generating such a function-level patch is nontrivial even for this simple example.
Since each row of the table is modified, the patch must involve a loop over the rows of the table. 
Synthesizing loops is challenging, 
because the synthesis algorithm must generate an inductive loop invariant.
Note, that the original data invariant is not suitable because it does not hold on entry to the new loop -- the programmer's loop broke the data invariant in
the first place.
Moreover, even if the synthesis algorithm is clever enough to generate a loop, it must be careful to
preserve the programmer's original logic. The  simplest solution is to update the \T{daily} field of each
row using the \T{weekly} values. Such a patch would be disastrous -- the data invariant
is erroneously ``maintained'' by undoing the programmer's changes!

More generally, this simple example highlights the two main research problems for synthesis-based language support of data invariants:
\begin{inparaenum}[(a)]
\item \emph{complex patches:} even for simple data invariants, the synthesis algorithm must calculate both inductive 
invariants and complex control flow, and 
\item \emph{the frame problem:} without frame conditions, the synthesis algorithm can enforce the invariant by simply reverting the programmer's changes.
\end{inparaenum}

\subsection{Targeted Synthesis and the Spyder Language}

The technical contribution of this paper is a solution to the above two research problems. 
Our solution consists of co-designing a programming language 
with a novel \emph{targeted synthesis} algorithm,
which generates patches \emph{locally} 
-- as close as possible to the invariant violation --
as opposed to at the function boundaries.

Targeted synthesis addresses the problem of \emph{complex patches}
by generating multiple patches that are as local as possible.
For example, in \autoref{fig:budget-code}, 
a local patch updates \T{r.week} and \T{r.total} inside the loop.
Local patches are typically much smaller;
moreover, pushing a patch inside a loop 
often results in preserving the original data invariant between loop iterations,
creating an inductive loop invariant. 
In our example, not only is the desired patch a short, straight-line code snippet,
but also it maintains data invariant (1) as an inductive loop invariant.

Targeted synthesis also addresses \emph{the frame problem}:
enforcing invariants at basic block boundaries enables a simple syntactic check
that disallows patching variables modified by the programmer in that block
and thereby ensures that all programmer's changes are preserved.

This paper presents \tool,
a core language with iterators and data invariants,
which is designed to be amenable to targeted synthesis.
In particular, \tool offers iterator-based loops and iterator-based data invariants,
which allows the synthesis algorithm to exploit their structural similarity
and push synthesis specifications inside loops,
in order to generate local patches. 




The remainder of the paper is structured as follows. 
We use the domain of web GUI applications to give a high-level overview of \tool in \autoref{sec:overview}. 
\autoref{sec:language} formalizes the semantics of the \tool language,
and \autoref{sec:synthesis} presents our targeted synthesis algorithm for extending \tool programs with invariant-preserving patches.
As part of our formalisms, we contribute a soundness guarantee that the targeted synthesis algorithm preserves the original invariants; this is summarized in \autoref{sec:synthesis}.
\autoref{sec:evaluation} evaluates our \tool compiler on a series of benchmark and case studies.
Finally, we conclude by reviewing related work in \autoref{sec:related}.

\subsection{Main Contributions}
The main contributions of this paper are:
\begin{compactenum}[~1)]
\item Programming with data invariants:
a new programming model, where the developer explicitly states relational data invariants,
and a synthesis engine automatically generates code patches to maintain these invariants.
\item \technique: a sound and efficient algorithm for synthesizing patches for 
the restricted but useful class of data invariants we call iterator-based invariants.
\item \tool, a prototype implementation of \technique;
our empirical evaluation shows that \tool programs are concise and compositional,
and that \technique generates patches more efficiently than traditional program synthesis techniques. 
\end{compactenum}

%% file: overview.tex

\section{Overview}\label{sec:overview}

We begin with an overview of \technique on the budgeting application shown in \autoref{fig:budget},
in which the programmer uses \emph{data invariants} to author an interactive GUI application.
The rendering and logic of the application
are relatively easy to express using the imperative, as we will discuss
in \autoref{sec:related}, but this approach does not offer language support for statically enforcing application \emph{data invariants}. 
We will demonstrate how \tool supports data invariants by iteratively building the interactive logic for this example.

\subsection{Data Invariants}

\lstset{basicstyle=\lstfigstyle}

\begin{figure}[h]
  \centering
    \begin{subfigure}{.45\linewidth}
      \begin{lstlisting}[numbers=left]
data weeks: int[];
data days: int[];

@\cbone{foreach w in weeks, d in days:}@:
  @\cbone{7 * d.val = w.val}@

procedure adjustForCOLA(cola: int):
  for d in days:
    if (d > 0):
      d <- d * cola;
\end{lstlisting}
    \caption{Source code in the \tool language for the days and weeks columns of the budgeting application. }
    \label{subfig:budget-source-ex1}
    \end{subfigure}
    \begin{subfigure}{.45\linewidth}
      \begin{lstlisting}
// procedure adjustForCOLA(cola: int):
for d in days, @\cbone{w in weeks}@:
  if (d > 0):
    d <- d * cola;
    @\cbone{w <- 7 * d;}@
\end{lstlisting}
    \caption{Generated \tool code for adjustForCOLA in the budgeting application.}
    \label{subfig:budget-generated-ex1}
    \end{subfigure}
  \caption{Programming with an invariant between Days and Weeks in the budgeting application.}
  \label{fig:budget-spyder-ex1}
\end{figure}

\lstset{basicstyle=\lstmainstyle}
The programmer starts with the logic for the Weekly and Daily columns, shown in \autoref{fig:budget-spyder-ex1}. To do this, the programmer
declares a collection of \T{int}s termed \T{weeks}, shown on line 1 of \autoref{subfig:budget-source-ex1}, as well as a collection of \T{int}s termed \T{days} (line 2). 
These two declarations introduce new mutable global variables \T{days} and \T{weeks}. 

One invariant of the system is the \emph{unit-conversion} invariant (invariant (\ref{unitconv-invariant}) in \autoref{sec:motivating-example}): each of the elements of \T{weeks} is $7$ times greater than the corresponding
element of \T{days}. This invariant should always hold and in particular, needs to be enforced whenever either \T{weeks} or \T{days} is mutated.
To specify the unit-conversion invariant, the programmer uses a \T{foreach} construct
on line 4, binding the elements of \T{weeks} to the local iterator \T{w} and the elements of \T{days} to \T{d}. Using these local bindings, they express the unit-conversion invariant using the formula
on line 5: \T{7 * d.val = w.val}. 

Because this unit-conversion invariant is defined over elements of collections, traditional techniques
would model collections as arrays and require a \emph{quantified relation} over the indicies of the arrays. Such relations
are notoriously tricky to build by hand (and indeed, to verify), but in \tool, the programmer
can use the \T{foreach} abstraction. This abstraction builds an 
element-wise product relation by introducing fresh \emph{iterator} bindings over the abstracted collections. 

\subsection{Maintaining Data Invariants with Spyder}
Next, the programmer writes imparative code implementing the desired functionality, without correcting for the violated unit-conversion invariant, as \tool will patch to maintain it.
In the application, recall that the budget-builder needs to adjust all of the revenues (and only the revenues) in the budget by the Cost-of-Living-Adjustment (COLA). To implement this modification, the 
programmer writes a \emph{procedure} called \T{adjustForCOLA} on line 7. This function iterates over the elements of \T{days} using the \T{for} loop on line 8, which
binds each element of \T{days} to a local iterator variable \T{d}. 

Since the COLA should only be applied to revenues, the programmer checks the value of the element \T{d} using a conditional on line 9, and then scales the 
daily revenue by an iterator update on line 10. The iterator semantics of \tool are standard for object-oriented iterators; in particular, notice that the value of the iterator (e.g. \T{d.val}) is implicitly given
by the iterator variable itself (e.g. \T{d} in the expression \T{d > 0}).


In this code snippet, the programmer has directly assigned an updated value to \T{d}, and by extension the values of \T{days}.
On its own, this update breaks the unit-conversion invariant -- in particular, the Weekly value of this row of the application depends on the concrete value of \T{d}. Using traditional
techniques, the programmer would have to manually maintain the invariant by setting
the corresponding value of \T{weeks}, i.e. by adding an extra snippet for correctly updating \T{weeks}. 

Fortunately for the programmer, invariants are statically maintained in \tool and the compiler
synthesizes and inserts a invariant-restoring snippet automatically, as shown in \autoref{subfig:budget-generated-ex1}. In this case,
the compiler \emph{extends} the original loop over \T{days} with an extra binding over \T{weeks}; in \tool, this has the semantics of 
a simultaneous iteration (analogous to a functional zip) so that \T{d} and \T{w} refer to elements of \T{days} and \T{weeks} at the same
index.

More generally, in contrast to 
traditional programming, \tool enables the programmer to write modifications that are
\emph{agnostic} to the existing
invariants. In this case, the programmer simply writes a direct update to the elements of \T{days} and \tool ensures that the overall system's state is correct.  

\subsection{Program Composition through Data Invariants}

In this subsection we demonstrate code evolution with \tool.
At some later date, the programmer adds a feature to the budget application: a running totals column to help track the state of the budget.
To do this, they add a collection for the total values, and the data invriant to populate it, seen in \autoref{subfig:budget-source-ex2}, lines 8-9.
In order to define the running-sum property, \tool provides an iterator method called \T{prev}, which allows access to the previous value of the iterator. This is useful for defining accumulator properties or enforcing sortedness. 
\tool will generate the implementation of populating the totals column in its entirety.

However, since time has passed since the last change made to the system, the programmer has forgotten about \T{adjustForCOLA}, which breaks our new totals invariant.
In a traditional imparative programming paradigm, it would be the programmer's responsibility to track down every function that breaks the invariant and fix it.
However, with \tool, the compiler checks the new invariant against all existing functions and generates a new patch to \T{adjustForCOLA} to ensure it is maintained.

The different invariants are compositional from the user's perspective---in practice, each function is checked against all invariants in the code. It is the responsibility of \technique to find in a failed set if invariants the actual invariants that have failed, and to reduce those to a local specification that can be used to synthesize a patch. This is shown in \autoref{sec:synth:algo}.

In evolving the codebase, the programmer later adds another feature, coloring negative values in red.
This is done using two sets of invariants: one for totals (lines $12$-$14$), and one for days (lines $16$-$19$).
Notice that \T{adjustForCOLA} does not invalidate the days invariant.
\tool checks this in compile time, resulting in no changes being made to \T{adjustForCOLA}---%
as opposed to dynamic techniques which would generate code that tests this in runtime.

\lstset{basicstyle=\lstfigstyle}
\begin{figure}[t]
  \centering
    \begin{subfigure}{.45\linewidth}
      \begin{lstlisting}[numbers=left]
data weeks: int[];
data days: int[];

@\cbone{foreach w in weeks, d in days:}@
  @\cbone{7 * d.val = w.val}@

data totals: int[];
@\cbtwo{foreach d in days, t in totals:}@
  @\cbtwo{t.val = t.prev(0) + d.val}@

data totalFontColors: int[];
@\cbthr{foreach t in totals, c in totalFontColors:}@
  @\cbthr{(t.val >= 0 <=> c.val = black)) \&\&}@
  @\cbthr{(t.val <  0 <=> c.val = red)}@

data rowFontColors : int[];
@\cbthr{foreach d in days, c in rowFontColors:}@
  @\cbthr{(d.val >= 0 <=> c.val = black)) \&\&}@
  @\cbthr{(d.val <  0 <=> c.val = red)}@

\end{lstlisting}
    \caption{\tool source code for an accumulated Totals invariant. }
    \label{subfig:budget-source-ex2}
    \end{subfigure}
    \begin{subfigure}{.45\linewidth}
      \begin{lstlisting}
// procedure adjustForCOLA(cola: int):
for d in days, @\cbone{w in weeks}@, @\cbtwo{t in totals}@,  @\cbthr{cr in rowFontColors, ct in totalFontColors}@:
  @\cbtwo{t <- t.prev(0) + d;}@
  @\cbthr{if (t < 0):}@
    @\cbthr{ct <- red;}@
  @\cbthr{else:}@
    @\cbthr{ct <- black;}@
  if (d > 0):
    d <- d * cola;
    @\cbone{w <- 7 * d;}@
    @\cbtwo{t <- t.prev(0) + d;}@
    @\cbthr{if (d < 0):}@
      @\cbthr{cr <- red;}@
    @\cbthr{else:}@
      @\cbthr{cr <- black;}@
    @\cbthr{if (t < 0):}@
      @\cbthr{ct <- red;}@
    @\cbthr{else:}@
      @\cbthr{ct <- black;}@
\end{lstlisting}
    \caption{Generated \tool code for an update to Days.}
    \label{subfig:budget-generated-ex2}
    \end{subfigure}
  \caption{Programming with an accumulator invariant between \T{days} and \T{totals}, as well as a font color invariant. Colors indicate the relationship between the invariant (a) and generated code (b).}
  \label{fig:budget-spyder-ex2}
\end{figure}
\lstset{basicstyle=\lstmainstyle}

    
The code generated by \tool is seen in \autoref{subfig:budget-generated-ex2}. 
The fixes introduced by \tool are nontrivial in several ways: 
\begin{inparaenum}[(1)]
\item the fixes are \emph{extensive}, accounting for the majority of the code in \autoref{subfig:budget-generated-ex2},
\item the fixes are \emph{nonlocal}, meaning that each fix is spread out over (and interleaved with) the original code,
\item the fixes have to \emph{add new variables} just to maintain the invariants, and
\item each invariant requires multiple fixes.
\end{inparaenum}

\subsection{Generalization of Technique}
From the programmer's perspective, the process of invariant patching is invisible -- \tool accepts the original, invariant-oblivious code. More generally, the cognitive load of imperative programming with invariants in \tool is significantly less than 
traditional techniques. Using \technique, when writing imperative code, the programmer needs to only reason about \emph{local} code properties (e.g. the value of \T{d}) and does not need to reason about \emph{global} code invariants (e.g. the relation between \T{days} and \T{rowFontColors}).

The structure of the remainder of this paper is as follows:
We first describe the \tool source language in \autoref{sec:language}, and show how to translate
from \tool terms to a well-studied imperative language.
We also give a hoare-style axiomatic semantics to \tool programs, and provide a formal guarantee that \tool verification triples are equivalent to standard triples.
In \autoref{sec:synthesis}, we present synthesis rules for patching and extending \tool programs, and provide a formal guarantee that our synthesis rules are sound with respect to our \tool verification triples.
Finally, we present several case studies and a benchmarking evaluation in \autoref{sec:evaluation}, and conclude with a discussion of related work in \autoref{sec:related}.


%% file: language.tex
\section{The Spyder Language}\label{sec:language}
We present the syntax and semantics of the \tool source language. In this section, we do not describe how to maintain data invariants. 
Instead, we just provide the formal framework for both expressing collection-manipulation programs, as well as axiomatically verifying properties over programs.
We will build on the results of this section in \autoref{sec:synthesis} to show how to maintain data invariants through \technique.

First, we introduce the core syntax of \tool in \autoref{subsec:syntax}. Then, we give a semantics to the syntax by translating \tool terms to a well-studied standard imperative language in \autoref{subsec:translation}.
Next, we demonstrate how to mechanically verify when invariants are maintained or violated by defining a Hoare-style \cite{hoare1969axiomatic,floyd1993assigning} axiomatic logic for \tool in 
\autoref{subsec:verif}.
Finally, we give a proof of soundness for this verification logic by reduction to the standard axiomatic semantics for imperative array programs (i.e. Hoare logic) in \autoref{thm:rel-sound}.

\subsection{Surface Syntax for Spyder}\label{subsec:syntax}
At its core, \tool is an imperative collection-manipulation language. The focus in \tool is to support data invariants for 
mutable, finite collections. To this end, we formalize and define a core calculus for iterating over and mutating collections, which we present in \autoref{subfig:spy-syntax}. 
\paragraph{Values and Types} \tool has three datatypes: \emph{integers}, \emph{collections}, and \emph{iterators}. 
\emph{Integers} are standard and we denote a variable declaration of type
int as \T{data x: int}.

\paragraph{Collections}  Collections hold elements and are analogous to ordered containers, e.g. lists or arrays. 
For variable declarations, we denote a collection of \T{T} by \T{data col: T[]}. Collections are homogeneous and for clarity of presentation, our core syntax and formalisms assume that all collections are 1-dimensional (i.e. collections of integers). 
In our implementation, however, collections can nest arbitrarily (and extending the formalisms to arbitrary nesting is straightforward). For example, the list \T{[1,2,3]} is a collection of integers and the list \T{[[1,2],[3,4]]} is a collection of integer collections. In contrast, the list \T{[1,[2,3]]} has mixed element types and is not valid.
Collections expose a single method, $\concsize$, which returns the number of elements in the collection. For simplicity, we assume all collections
have a statically known size which does not vary at runtime. 
We also assume that collection sizes are homogeneous, for example, the list \T{[[1,2],[3,4,5]]} would not be a valid \tool collection. 
A key difference between collections and arrays is that collections do not support subscription (i.e. \T{col[idx]} is not a valid \tool term).
Instead, to access the elements of a collection, \tool exposes the $\concfor \, \pair{\sx}{\sy}$ statement, which iterates over the values of the collection $\sy$.
In addition to iteration, the $\concfor$ statement creates a new variable binding for an \emph{iterator} variable.

\paragraph{Iterators} Iterators allow access to the underlying elements of a collection. Iterator variables are not explicitly declared using \T{data} but are instead created in $\concfor$ loops. 
\tool supports several standard iterator methods: $\concelem$, which returns the value of the iterated collection; $\concidx$, which returns the current iteration index; $\concprev$, which returns
the previous value; and the $\sx \Put E$ operator (termed ``put''), which destructively \emph{updates} the value of the iterator $\sx$ with the expression $E$. 
For example, after the evaluation of the term \T{for x in xs: x <- x + 1;}
each element of \texttt{xs} is incremented by exactly one.

\paragraph{Statements}
For control-flow, \tool has mostly standard imperative statements. A key exception is the $\concfor$ term, which as discussed above, iterates over a collection.
In addition, the $\concfor$ loop iterates over multiple collections simultaneously, similar to a zip in function programming. For example the term 
\T{for x in xs, y in ys: x <- x.val + 1; y <- y.val + 1;} replaces each element in \T{y} with the corresponding element in \T{x}. Furthermore, iteration is only well-defined when the iterated collections have the same size. 

\paragraph{Specifications}
To express specifications for \tool terms, \tool exposes a rich specification language, the $\sInv$ terms. To ease the burden of synthesis and 
verification, we syntactically phrase specifications in a conjunctive normal form.
At the top level are conjunctions of specifications using the $\land$ operator.
Each conjunct can be either a bare expression, or a quantification term. For quantification,
\tool supports two quantifiers: 
\begin{inparaenum}[(1)]
  \item An existential quantifier through the $\concexists$ keyword. This quantifier is not present in the surface syntax of \tool and is only used in the axiomatic semantics, which we present in \autoref{subsec:verif}.
  \item A universal quantifier through the $\concforeach$ keyword, which quantifies over the elements of a collection. For example, the specification \T{foreach x in xs, y in ys: x.val > y.val} states that each element of \T{xs} is greater than the corresponding element of \T{ys}. Similar to the $\concfor$ statement, the $\concforeach$ term is only well-defined when the bound collections have the same size. We discuss the details of specifications more in \autoref{subsec:verif}.

\end{inparaenum}

\begin{figure}
  \centering
  \scriptsize
  \begin{subfigure}[b]{0.5\linewidth}
  \[
  \begin{array}{lrl}
    &         & \sv, \su \in \sVars, \quad \si \in Z \\
        \sInv   & \bnfdef & \concforeach \; \; \set{\pair{\svi}{\sui}} \; \; \sInv \\
                & \bnfalt & \concexists \; \sv \; . \; \sInv \\
                & \bnfalt & \sInv \; \Invand \; \sInv \\
                & \bnfalt & \sExpr \\
        \sBlock & \bnfdef & \blockbase \bnfalt \sStmt \; \blocksep \; \sBlock \\ 
        \sStmt  & \bnfdef & \sv \; \Assign \; \sExpr \\
                & \bnfalt & \sv \; \Put \; \sExpr \\
                & \bnfalt & \cif \; \sExpr \; \cthen \; \sBlock \; \celse \; \sBlock \\ 
                & \bnfalt & \concfor \; \; \set{\pair{\svi}{\sui}} \; \; \sBlock \\ 
        \sExpr  & \bnfdef & \sv \bnfalt \si \bnfalt \conctrue \bnfalt \concfalse \\
                & \bnfalt & \sExpr \; \sbop \; \sExpr \\
                & \bnfalt & \suop \; \sExpr\\
                & \bnfalt & \sv . \concelem\\
                & \bnfalt & \sv . \concprev(\sExpr) \\
                & \bnfalt & \sv . \concidx \\ 
                & \bnfalt & \sv . \concsize \\
        \sbop   & \bnfdef & \plus \bnfalt \times \bnfalt \% \bnfalt \implies \bnfalt \Longleftrightarrow \bnfalt \ldots \\
        \suop   & \bnfdef & \lnot \bnfalt ! \\
  \end{array}
  \]
  \caption{Syntax for the \tool language.}
  \label{subfig:spy-syntax}
  \end{subfigure}%
  ~
  \begin{subfigure}[b]{0.5\linewidth}
    \[
      \begin{array}{lrl}
                  &         & \boogv, \boogu \in \boogVars, \quad \boogi \in \mathbb{Z} \\
        \boogStmt  & \bnfdef & \boogv \; \Assign \; \boogExpr \\
                & \bnfalt & \boogv \; [ \; \boogExpr \; ] \; \Assign \; \boogExpr \\
                & \bnfalt & \cif \; \boogExpr \; \cthen \; \boogStmt \; \celse \; \boogStmt \\ 
                & \bnfalt & \concwhile \; \; \boogExpr \; \boogStmt \\ 
                & \bnfalt & \boogStmt \; \blocksep \; \boogStmt \\ 
                & \bnfalt & \blockbase \\
        \boogExpr  & \bnfdef & \boogv \bnfalt \boogi \bnfalt \conctrue \bnfalt \concfalse \\
                & \bnfalt & \boogExpr \; \boogbop \; \boogExpr \\
                & \bnfalt & \cif \; \boogExpr \; \cthen \; \boogExpr \; \celse \; \boogExpr \\
                & \bnfalt & \booguop \; \boogExpr\\
                & \bnfalt & \boogExpr \; [ \; \boogExpr \; ] \\
                & \bnfalt & \concsize (\boogv)\\
                & \bnfalt & \forall \; \boogv \; . \;  \boogExpr \\
                & \bnfalt & \exists \; \boogv \; . \;  \boogExpr \\
        \boogbop   & \bnfdef & \plus \bnfalt \times \bnfalt \% \bnfalt \implies \bnfalt \Longleftrightarrow \bnfalt \ldots \\
        \booguop   & \bnfdef & \lnot \bnfalt ! \\
      \end{array}
    \]
    \caption{Syntax for the \targetLang language.}
    \label{subfig:imp-syntax}
  
  \end{subfigure}

  \caption{Syntax for \tool and \targetLang.}
  \label{fig:syntax}

\end{figure}

\subsection{Imperative Target Language}\label{subsec:translation}

We formalize the semantics of \tool by translating to an idealized imperative verification language, which we call \targetLang.
The syntax of this verification target language is shown in \autoref{subfig:imp-syntax}.
This language is very similar to Boogie~\cite{leino2008boogie} and indeed, in our implementation, we compile and synthesize to Boogie.

Although \tool and \targetLang have similar syntax, there are 
several major differences. Broadly speaking, \targetLang does not have language support for either collections or iterators.
\targetLang instead offers mutable low-level arrays, which map (integer) indices to values. At the statement level \targetLang
supports mutable updates to both variables and arrays, as well as general $\concwhile$ loops. For expressions, \targetLang enables
rich quantification through the $\forall$ quantifier, but in contrast to \tool, does not support iterator methods.

\lstset{basicstyle=\lstfigstyle}
\begin{figure}
  \centering
  \small
    \begin{subfigure}{.45\linewidth}
      \begin{lstlisting}[numbers=left]
data values: int[];
data product: int[];

foreach v in values, fact in product:
  fact.val = fact.prev(1) * v.val

procedure multValues():
  for v in values, fact in product:
    v <- v.val * 1.05;
    fact <- fact.prev(1) * v.val;

\end{lstlisting}
    \caption{\tool source code for a product invariant. }
    \label{subfig:lang-prod-src}
    \end{subfigure}
    \begin{subfigure}{.45\linewidth}
      \begin{lstlisting}

var values: [int]int;
var v: int;
var product: [int]int;
var fact: int;

procedure multValues(){
  v := 0; fact := 0;
  while 
    (v < size(values) && fact < size(product))
// invariant: 
// forall i. 0 <= i < v ==> 
//    product[i] == values[i] * 
//      (if i == 0 then 1 else product[i-1])
  {
    val[v] := val[v] * 1.05;
    if (index == 0) {
      product[fact] := val[v];
    } else {
      product[fact] := product[fact-1] * val[v];
    }
    fact := fact + 1; v := v + 1;
  }
}
\end{lstlisting}
    \caption{Translated \targetLang code for a product invariant.}
    \label{subfig:lang-prod-target}
    \end{subfigure}
  \caption{Source code and translation maintaining a product invariant. 
  In contrast to the examples in \autoref{sec:overview}, the source code maintains the invariant, and the translation step must soundly produce ImpArray code which also maintains it.  }
  \label{fig:lang-prod}
\end{figure}
\lstset{basicstyle=\lstmainstyle}

To support collections and iterators, the translation from \tool to \targetLang must implement collection and iterator
logic in terms of arrays and indices. We show an example of this in \autoref{fig:lang-prod}, in which a \tool program for calculating a product
is translated into \targetLang. In this case, the integer collections \T{values} and \T{product} in \tool map 1-to-1 to arrays in \targetLang, and the 
\T{for} loop in \tool is desugared into a \T{while} loop with an explicit index in \targetLang.
At a high-level, collections in \tool correspond 1-to-1 with arrays in \targetLang, and iterator variables in \tool
correspond to indices in \targetLang.
This example is similar to a subproblem of the \technique algorithm (discussed in detail in \autoref{sec:synthesis}), which 
reasons about candidate programs like multValues.

\subsection{Overview of Translation Semantics}

We formalize translation as a syntax-directed recursive function over \tool terms given in \autoref{fig:translation}. 
Since \T{for} loops bind iterator variables, the translation must be stateful. We choose to explicitly pass the state using finite
mathematical maps, which we term \emph{translation contexts} and we generally denote as $\Gamma$. We denote the translation of a term $t$
using the context $\Gamma$ as the \targetLang term $\translate{t}{\Gamma}$; we refer to this as ``the translation of $t$ in the context of $\Gamma$''.

\begin{figure}[h]
  \centering
  \scriptsize
  \begin{subfigure}{\linewidth}
    \[  
      \begin{array}{lcl}
        \text{\tool Term} & & \text{\targetLang Term} \\ \hline
        \translate{\sv}{\Gamma} & = & \boogv \\
        \translate{\si}{\Gamma} & = & \boogi \\
        \translate{\conctrue}{\Gamma} & = & \conctrue \\
        \translate{\concfalse}{\Gamma} & = & \concfalse \\
        \translate{E_l \; \sbop \; E_r}{\Gamma} & = & \translate{E_l}{\Gamma} \; \boogbop \; \translate{E_r}{\Gamma} \\
        \translate{\suop \; E}{\Gamma} & = & \booguop \; \translate{E}{\Gamma} \\
        \translate{\sv . \concelem}{\Gamma} & = & \Gamma(\sv) [\boogv] \\
        \translate{\sv . \concprev(E)}{\Gamma} & = & \Cond{\boogv > 0}{\Gamma(\sv)[\boogv - 1]}{\translate{E}{\Gamma}} \\
        \translate{\sv . \concidx}{\Gamma} & = & \boogv \\
        \translate{\sv . \concsize}{\Gamma} & = & \concsize(\sv) \\
      \end{array}  
    \]
    \caption{Translation rules for Spyder Expressions to ImpArray Epressions.}
  \end{subfigure}
  \begin{subfigure}{\linewidth}
    \[
      \begin{array}{lcl}
        \text{\tool Term} & & \text{\targetLang Term} \\ \hline
        \translate{\concforeach \; \pair{\sv}{\su} \; I}{\Gamma} & = & \forall \; \boogv \; . \; (0 \leq \boogv \land \boogv < \concsize(\boogu)) \implies \translate{I}{\Gamma \oplus \sv \mapsto \boogu } \\
        \translate{\concexists \; \sv \; I}{\Gamma} & = & \exists \; \boogv \; . \; \translate{I}{\Gamma} \\
        \translate{I_l \land I_r}{\Gamma} & = & \translate{I_l}{\Gamma} \land \translate{I_r}{\Gamma} \\
      \end{array}  
    \]
    \caption{Translation rules for Spyder Specifications to ImpArray Epressions.}
  \end{subfigure}
  \begin{subfigure}{\linewidth}
    \[
      \begin{array}{lcl}
        \text{\tool Term} & & \text{\targetLang Term} \\ \hline
        \translate{\sv \Assign E}{\Gamma} & = & \boogv \; \Assign \; \translate{E}{\Gamma} \\
        \translate{\Cond{E}{B_t}{B_f}}{\Gamma} & = & \Cond{\translate{E}{\Gamma}}{\translate{B_t}{\Gamma}}{\translate{B_f}{\Gamma}} \\
        \translate{\sv \Put E}{\Gamma} & = & \Gamma(\sv) [\boogv] \; \Assign \; \translate{E}{\Gamma} \\
        \translate{\concfor \; \pair{\sx}{\sy} B_i}{\Gamma}  & = & \boogx \; \Assign \; 0 \; \blocksep \\
                                                             &   & \; \concwhile \; (\boogx < \concsize(\boogy)) \; \translate{B_i}{\Gamma \oplus \sx \mapsto \boogy} \; \blocksep \boogx \; \Assign \; \boogx + 1 \\
        \end{array} 
    \]
    \caption{Translation rules for Spyder Statements to ImpArray Statements.}
  \end{subfigure}
  \caption{Translation rules for Spyder to ImpArray}
  \label{fig:translation}
\end{figure}


\paragraph{Well-formedness of Translation Contexts} In general, the translation process is only well-defined if the translation context $\Gamma$
is well-formed. Intuitively, there must be no name-collisions; a collection must not be iterated over multiple times; an iterator variable must not 
be directly written to (i.e. using assignment $\Assign$ instead of the iterator $\Put$ operator); etc. We formalize these well-formedness constraints 
in \autoref{fig:trans-wellform}, which relates \tool terms $t$ to translation contexts that are well-formed for translating $t$. We denote
a well-formed term using $\wellform{t}{\Gamma}$ and we say $\Gamma$ is well-formed with respect to $t$.

\begin{figure}
  \centering
  \scriptsize
  \begin{subfigure}{\linewidth}
    \[
      \begin{stackCC}
      \inference[Var-Global]{
        \sv \in \globals \\
        \sv \notin \Gamma
      }{
        \wellform{\sv}{\Gamma}
      } \quad
      \inference[Var-Bound]{
        \sv \notin \globals \\
        \sv \in \Gamma
      }{
        \wellform{\sv}{\Gamma}
      } \\ \\ 
      \inference[Prim-Int]{
      }{
        \wellform{\si}{\Gamma}
      } \quad
      \inference[Prim-BT]{
      }{
        \wellform{\conctrue}{\Gamma}
      } \quad
      \inference[Prim-BF]{
      }{
        \wellform{\concfalse}{\Gamma}
      } \\ \\

      \inference[Bop] {
        \wellform{E_l}{\Gamma} \quad \wellform{E_r}{\Gamma}
      } {
        \wellform{E_l \; \sbop \; E_r}{\Gamma}
      } \quad
      \inference[Uop] {
        \wellform{E}{\Gamma}
      } {
        \wellform{\suop \; E}{\Gamma}
      }  \quad

      \inference[Elem] {
        \sv \in \Gamma
      } {
        \wellform{\sv . \concelem}{\Gamma}
      } \\ \\
      \inference[Prev] {
        \sv \in \Gamma \quad \wellform{E}{\Gamma}
      } {
        \wellform{\sv . \concprev(E)}{\Gamma}
      } \quad
      \inference[Idx] {
        \sv \in \Gamma
      } {
        \wellform{\sv . \concidx}{\Gamma}
      } \quad
      \inference[Size] {
        \sv \in \range{\Gamma}
      } {
        \wellform{\sv . \concsize}{\Gamma}
      } 

    \end{stackCC}
    \]
    \caption{Well-formedness rules for Spyder Expressions.}
  \end{subfigure}
  \begin{subfigure}{\linewidth}
    \[
      \begin{stackCC}
        \inference[Foreach] {
          \su \notin \Gamma \quad \su \notin \range{\Gamma} \quad \sv \notin \Gamma \quad \su \in \globals \\
          \wellform{I}{\Gamma \oplus \sv \mapsto \su}
        } {
          \wellform{\concforeach \, \pair{\sv}{\su} I}{\Gamma}
        }\\
        \inference[Exists] {
          \fresh{\sx} \quad \wellform{I}{\Gamma}
        } {
          \wellform{\concexists \; \sx \; . \; I}{\Gamma}
        }\quad 
        \inference[Conjunct] {
          \wellform{I_l}{\Gamma} \\
          \wellform{I_r}{\Gamma} 
        } {
          \wellform{I_l \land I_r}{\Gamma}
        }
        
      \end{stackCC}  
    \]
    \caption{Well-formedness rules for Spyder Invariants.}
  \end{subfigure}
  \begin{subfigure}{\linewidth}
    \[ 
    \begin{stackCC}
      \inference[Blk-Skip]{
      }{
        \wellform{\blockbase}{\Gamma}
      }\quad
      \inference[Blk-Seq]{
        \wellform{S}{\Gamma} \quad \wellform{B}{\Gamma}
      }{
        \wellform{S \; \blocksep \; B}{\Gamma}
      }\\ \\     
      \inference[Stmt-Assign]{
        \sv \notin \Gamma \quad
        \wellform{E}{\Gamma}
      }{
        \wellform{\sv \Assign E}{\Gamma}
      }\quad
      \inference[Stmt-Put]{
        \sv \in \Gamma \quad
        \wellform{E}{\Gamma}
      }{
        \wellform{\sv \Put E}{\Gamma}
      } \\ \\
      \inference[Stmt-Cond]{
        \wellform{E}{\Gamma} \quad \wellform{B_t}{\Gamma} \quad \wellform{B_f}{\Gamma}
      }{
        \wellform{\Cond{E}{B_t}{B_f}}{\Gamma}
      }\\ \\
      \inference[Stmt-For]{
        \sy \notin \range{\Gamma} \quad \sx \notin \Assigned{B_i} \quad \sy \in \globals \\
        \wellform{B_i}{\Gamma \oplus \sx \mapsto \sy} 
      }{
        \wellform{\concfor \; \pair{\sx}{\sy} B_i}{\Gamma}
      }

    \end{stackCC}
    \]
    \caption{Well-formedness rules for Spyder Statements.}
  \end{subfigure}
  \caption{Well-formedness rules for Spyder. For exposition, when rules bind a variable (e.g. $\concfor$) we only formalize the well-formedness for a single binding. The extension to multiple bindings is straightforward.}
  \label{fig:trans-wellform}
\end{figure}

\paragraph{Semantics for Spyder terms}
Since $\targetLang$ is well-studied and has a well-understood semantics (replicated in \autoref{app:imp-semantics}), 
we define the semantics of \tool by translating into \targetLang. For details, see \autoref{app:spy-semantics}. 

A keen observer will notice that \tool's semantics are focused on alias-free
iterator-based programs. $\targetLang$ has actually been studied in the context
of verifying more exotic language features, such as object-oriented invariants \cite{leino2004object}, 
concurrency \cite{cohen2009vcc}, general arrays \cite{leino2010dafny}, heap-manipulation \cite{pek2014natural}, etc. 
Because other systems have verified exotic programs
using the rich, low-level semantics of $\targetLang$, in the future 
the semantics of \tool can be extended to handle relational invariant maintenance
for more complicated languages.

\subsection{Verification in Spyder and ImpArray}\label{subsec:verif}
We next define and present an axiomatic semantics for \tool that \technique will use to mechanically verify
when invariants are preserved or violated by a statement.
In addition, we prove that \tool axiomatic semantics are sound with respect to the standard axiomatic semantics for \targetLang (i.e. Hoare triples).

\paragraph{Hoare Triples for ImpArray} We start by briefly reviewing axiomatic semantics in \targetLang which are well-known \cite{hoare1969axiomatic}.
The standard approach, called Hoare triples, are deduction rules for relating three terms: a precondition $P$, a statement $S$, and a postcondition $Q$, denoted by  $\hoarebr{P} \, S \, \hoarebr{Q}$. Intuitively, the rules derive a triple if and only if given the precondition $P$, the postcondition $Q$ holds after executing the statement $S$.  
We replicate these rules in \autoref{fig:imp-verification}. 

Notice that in standard axiomatic semantics, the loop rule requires an inductive invariant $I$ to be maintained on \emph{every} iteration. Furthermore,
the axiomatic rules do not contain a notion of termination. As a result, the triple $\hoarebr{P} \, S \, \hoarebr{Q}$ should only be interpreted as valid if the statement $S$ terminates. 
In our case, termination is orthogonal. Our well-formedness constraints ensure that all loops over finite collections terminate, and so in practice, this is not
an issue for our use of the axiomatic semantics of \targetLang.

\paragraph{Hoare Triples for Spyder} We next provide a similar axiomatic semantics for \tool terms. In this case, we derive a triple $\hoare{P} \; S \; \hoare{Q}$, 
which has the same intuitive interpretation, that given $P$, $Q$ holds after executing $S$. 
As part of our contribution, we prove that the logic of \autoref{fig:spy-verification} is \emph{relatively sound}: given a well-formed translation context,
the axiomatic rules are sound with respect to Hoare logic. Intuitively, if we prove a triple in the \tool semantics,
then the corresponding translated triple holds in Hoare's axiomatic semantics. More formally, let $P$ and $Q$ be \tool Expressions,
let $S$ be a \tool Statement, and let $\Gamma$ be a translation context. If $\Gamma$ is well-formed with respect to $P$, $Q$, and $S$,
and we derive the triple $\hoare{P} \, S \, \hoare{Q}$, then there exists a Hoare Triple for the corresponding translated terms in 
\targetLang:

\begin{theorem}[Relative Soundness]\label{thm:rel-sound}
  \begin{multline*}
    \forall P, \, S, \, Q, \, \Gamma \, . \, \wellform{P \land Q}{\Gamma} \land \wellform{S}{\Gamma}  \implies \\
      \hoare{P} \, S \, \hoare{Q} \, \implies \hoarebr{\translate{P}{\Gamma}} \, \translate{S}{\Gamma} \, \hoarebr{\translate{Q}{\Gamma}}
  \end{multline*}
\end{theorem}
We prove this property by induction over the derivation of the \tool Triple $\hoare{P} \, S \, \hoare{Q}$, given in \autoref{app:proof-axiomatic-sound}. The key parts of the proof are
the soundness of the {\tt Put} and {\tt For} rules which we discuss in detail below.

\begin{figure}
  \centering
  \scriptsize
    \[
  \begin{stackCC}
    \inference[Consequence]{
          P \implies P' \qquad Q' \implies Q \\
          \hoare{P'} \; S \; \hoare{Q'}
        }{
          \hoare{P} \; S \; \hoare{Q}
        } \,
        \inference[Conditional]{
          \hoare{P \land E} \;  B_t \; \hoare{Q} \\
          \hoare{P \land \lnot \; E } \; B_f \; \hoare{Q} \\ 
        }{
          \hoare{P} \; \Cond{E}{B_t}{B_f} \hoare{Q}
        }\\ \\
        \inference[Assign]{
          \fresh{v'} \\
        }{
          \hoare{P} \; v \Assign E \;  \hoare{\concexists \, v' \, . \, P[v \mapsto v'] \land v = E[v \mapsto v']} 
        }\quad
        \inference[Sequence]{
          \hoare{P} \; S \; \hoare{Q} \\
          \hoare{Q} \; B \; \hoare{R}
          }{
            \hoare{P} \; S \; \blocksep \; B \; \hoare{R} 
          }   \\ \\
        \inference[Put]{
          \fresh{v'} \\
        }{
          \hoare{P} \; v \Put E \; \hoare{\concexists \, v' \, . \, P[\concelem(v) \mapsto v'] \land \concelem(v) = E[\concelem(v) \mapsto v']} 
        }  \\ \\

        \inference[For]{
          \modify{B_i} \cap \free{I} = \emptyset \\
          \hoare{ \weakenprev{I} \land 0 \leq \concidx(\sx) < \concsize(\sy) } \; B_i \; \hoare{ I } \\
        }{
          \hoare{ \concforeach \, \pair{\sx}{\sy} \, I } \; \concfor \; \pair{\sx}{\sy} B_i \;  \hoare{ \concforeach \, \pair{\sx}{\sy} \, I}
        }\quad 
        \inference[Skip]{
        }{
          \hoare{P} \; \blockbase \; \hoare{P} 
        }

  \end{stackCC}
  \]
  \caption{Hoare-style verification logic for \tool. For exposition, we only formalize the relation loops with a binding. Since loops are only well-defined when the iterated collections have the same statically known size, the extension to multiple bindings is straightforward.}
  \label{fig:spy-verification}
\end{figure}

\paragraph{Strong Iterator Updates}
{\tt Put} is interesting because under the hood, the update $\sx \Put E$ translates to an array write (namely $\Gamma(\sx)[\boogx] \Assign E$).
This is potentially problematic because standard array semantics assume indices
can alias and so all information about the collection $\Gamma(\sx)$ is lost after the update. However, \tool has no variable aliasing.
Moreover, the well-formedness rules ensure that values of
the collection $\Gamma(\sx)$ can only be referenced through exactly one iterator $\sx$ and 
one expression $\sx . \concval$.\footnote{In particular the well-formedness relation prohibits a $\concforeach$ quantifier over a collection $\sy$ from entering the body of a loop over $\sy$.} Consequently, in the {\tt Put} rule we reason about the value of $\sx . \concval$ while soundly retaining information about the collection $\Gamma(\sx)$.

\paragraph{Quantifier introduction and maintenance}
A key requirement of the axiomatic semantics is to soundly reason about when loops maintain (or violate) universally quantified invariants (i.e. $\concforeach$ terms).
To that end, we provide a {\tt For} rule, which is similar to a standard $\concwhile$ 
rule in that the inductive invariant is on both sides of the statement. Unlike the Hoare $\concwhile$ rule, however, the {\tt For} rule for a loop \T{for x in xs} requires a top-level \T{foreach x in xs} as well.\footnote{If a top-level term is not in this form but is equivalent under renaming and quantifier shuffling, the {\tt Consequence} rule can be used to rewrite the term to make progress.}

In order to show that a \T{foreach} invariant is maintained by a \T{for} loop, it suffices to reason about each iteration of the loop in isolation. Due to the well-formedness constraints, the only way to modify the elements of a collection is through the $\Put$ operator. As a consequence the execution of a loop iteration cannot invalidate the results of previous iterations. Since the loop is guaranteed to execute for each element of the collection, the rule introduces a \T{foreach} quantifier after the loop is complete. 

Furthermore, it's tempting to assume the specialized invariant as a precondition to
verifying the loop body. If the invariant does not contain the $\concprev$ method, this is completely valid. 
However, the \T{prev} method complicates matters because each iteration does not necessarily establish \T{prev} for the next iteration. To address this situation,
we use the $\weakenprevname$ helper function to soundly weaken an expression with respect to \T{prev}.
As a result, the {\tt For} rule retains as much information as is soundly possible,
and enables automated verification and synthesis by removing a layer of quantification.

\subsection{Maintaining Data Invariants}
With an axiomatic semantics for \tool programs, 
we now consider several techniques for maintaining data invariants.
We use a simple midpoint program in \autoref{fig:midpoint}, in which two variables \T{l} and \T{r} sum to 10, to demonstrate these techniques.

\lstset{basicstyle=\lstfigstyle}
\begin{figure}[h]
  \centering
  \begin{subfigure}[t]{0.33\textwidth}
  \begin{lstlisting}
data l: int;
data r: int;

// invariant: l + r = 10

procedure incrL():
  l = l + 1;
  r = r - 1;
procedure incrR():
  r = r + 1;
  l = l - 1;
\end{lstlisting}
\caption{Imperative: program with no additional specifications.}
\label{subfig:midpoint-imp}
\end{subfigure}
\begin{subfigure}[t]{0.33\textwidth}
  \begin{lstlisting}
data l: int;
data r: int;

// invariant: l + r = 10
l = 10 - r
r = 10 - l

procedure incrL():
  l = l + 1;

procedure incrR():
  r = r + 1;

\end{lstlisting}
\caption{FRP: functional specifications for \T{l} and \T{r}.}
\label{subfig:midpoint-frp}
\end{subfigure}
\begin{subfigure}[t]{0.3\textwidth}
  \begin{lstlisting}
data l: int;
data r: int;

// invariant: l + r = 10

l + r = 10

procedure incrL():
  l = l + 1;

procedure incrR():
  r = r + 1;

\end{lstlisting}
\caption{\tool: a single relational specification for \T{l} and \T{r}.}
\label{subfig:midpoint-spy}
\end{subfigure}
\caption{Three different specification techniques used to implement a midpoint program in which \T{l} and \T{r} sum to 10. }
\label{fig:midpoint}
\end{figure}

\lstset{basicstyle=\lstmainstyle}

\paragraph{Imperative Invariant Maintenance}
The most common technique for invariant maintenance is to manually track invariants
and provide a patch that maintains the invariant. This is tedious and error prone
because the programmer must manually remember 
\begin{inparaenum}[1)]
\item what the invariant is, and
\item how to maintain the invariant when it breaks. 
\end{inparaenum}
For example, in the midpoint
program (\autoref{subfig:midpoint-imp}), the programmer must remember that
\T{l} must be decremented after \T{r} is incremented, and vice-versa. 

From the programmer's perspective, this is also the least compositional approach to invariant maintenance.
If the invariant changes, it is up to the programmer
to find all the patches and fix them.
However it is also the most performant technique; the runtime system simply executes the code.

\paragraph{Functional Invariant Maintenance}
An alternative approach to manual maintenance is the Functional-Reactive programming
(FRP) paradigm,
in which the programmer provides a functional specification for solving
the invariant, and the language runtime detects when the functional specification
should be invoked. 
In this example \autoref{subfig:midpoint-frp} the programmer
gives two functional specifications for \T{l} and \T{r}, each in terms of the other.
In return, the language runtime uses these specifications to perform invariant maintenance,
saving the programmer the need to reason about maintenance within the
implementation of \T{incrL()} or \T{incrR()}. The downside of this approach is that the
runtime system must dynamically track data-dependencies, 
incurring a runtime overhead compared to the imperative approach.
We discuss FRP in more detail in \autoref{sec:related}.

\paragraph{Spyder Invariant Maintenance}
Finally, \technique enables automatic \emph{relational} invariant
maintenance. In contrast to a functional specification, a \emph{relational} specification
does not easily admit a clear resolution for the specification. From the programmer's 
perspective, relational specifications are much more clear and concise. Consider in this example
the specification in \autoref{subfig:midpoint-spy}; it clearly and unambiguously captures the
data invariant that \T{l} and \T{r} sum to 10. 

The power and expressiveness of relational specifications comes at a cost.
One way to handle these rich relational invariants is to dynamically solve the relational
specification, similar to FRP. This incurs a significant runtime overhead (see \autoref{sec:related})
and moreover, when the specification is erroneous, dealing with the error falls to to the end-user of the code
and not the programmer.

Instead, we take the approach of solving these invariants at compile time using program synthesis.
In the next section, we detail exactly how \technique enables the programmer to use
relational specifications automatically within the \tool language.

%% file: synthesis.tex
\section{Targeted Synthesis for Spyder}\label{sec:synthesis}
In this section, we detail the automatic enforcing of data invariants.
We motivate and formalize the problem in \autoref{subsec:synth-problem},
then, in \autoref{sec:synth:algo} we present its solution in the \technique algorithm.
We prove the algorithm sound in \autoref{subsec:synth-soundness}.

Recall the budgeting example introduced in \autoref{sec:motivating-example}. 
\autoref{sec:overview} showed the specific case of the unit-conversion data invariant, which establishes the required relationship between daily and weekly values, seen in \autoref{subfig:budget-source-ex1}.
Throughout this section we will demonstrate our algorithm on this invariant.

\subsection{Automatic Enforcement of Data Invariants}\label{subsec:synth-problem}

Let $\precond$ be a $\sInv$ term, and $S$ be a \tool statement (i.e. a $\sStmt$ term). 
We say that $\precond$ is a \emph{data invariant} for $S$ if and only if $S$ maintains $\precond$:
~
\[
\hoare{\precond} \; S \; \hoare{\precond}.
\]
~
For example, the specification \T{foreach x in xs: x.val > 0} is a data invariant for a loop which increments each value of \T{xs}, \T{for x in xs: x <- x.val + 1}, but it is not a data invariant for decrement loop \T{for x in xs: x <- x.val - 1}.
This definition extend straightforwardly to statement blocks $B$.


Let $B, B'$ be two \tool blocks.
We say that a block $B'$ is an \emph{extension} of $B$ ($B\prec B'$) 
if $B$ and $B'$ have identical semantics on variables modified by $B$.



An \emph{invariant enforcement} problem is a pair $\langle B, \precond\rangle$ of a block $B$ and a specification $\precond$.
A solution to the enforcement problem is a block $B'$ such that $B\prec B'$ and $\hoare{\precond} \; B' \; \hoare{\precond}$.
In other words, the goal is to find an extension of $B$ such that $\precond$ is a data invariant for the extended block.

In our example, we wish the unit-conversion invariant on lines 4 and 5 to be a data invariant. This means the invariant enforcement problem is to enforce
this specification on the body of \T{adjustForCOLA}.

To find a solution,
our algorithm analyses $B$ and insert local patches whenever the invariant needs to be restored.
Since there are many candidate patches to explore, the key challenge is to make the search efficient.
To this end, our algorithm:
\begin{inparaenum}[(1)]
\item a-priori restricts the search to extensions of $B$,
by keeping track of the set of variables that a patch is allowed to modify;
\item \emph{targets} the invariant $\precond$ to $B$, 
producing a specification for each patch that is as local as possible. 
\end{inparaenum}

In this example, because \T{adjustForCOLA} modifies the elements of \T{days}, our algorithm
must find an extension that has an equivalent effect on \T{days}. Further, since 
\T{adjustForCOLA} iterates over \T{days}, our algorithm will \emph{target} the 
data invariant on lines 4 and 5 to a local specification, specific to just the loop
body on lines 9 and 10. We next explain the details of our algorithm.


\subsection{Targeted Synthesis Algorithm}\label{sec:synth:algo}

We formalize \technique as a \emph{completion judgment}
$\staleVS \; \vdash \; \hoare{\precond} \; B \; \hoare{\postcond} \; \redexstep \; B'$.
Intuitively, given a pre- and post-condition $\precond$ and $\postcond$,
and the set of variables $\staleVS$ modified so far,
an input block $B$ should be completed into $B'$.
In this case, we say that $B'$ is a \emph{completion} for $B$,
and the intension is that $B'$ satisfies the specification ($\hoare{\precond} \; B' \; \hoare{\postcond}$)
and does not modify any variables in $\staleVS$ (i.e. $\modify{B} \cap \staleVS = \emptyset$).
We present the inference rules for this judgment in \autoref{fig:synth-rules}.

\begin{figure}[ht]
  \centering
  \scriptsize
    \[
    \begin{stackCC}
      \inference[Synth-Base]{
        \metafont{cands} =  \{v \; \vert \; v \; \metarel_{\precond} \; y, \; y \in \staleVS \} \\
        \modify{B} \subseteq (\metafont{cands} \setminus \staleVS), \; \hoare{\precond} \; B \; \hoare{\postcond} \\
      }{
         \staleVS \; \vdash \; \hoare{\precond} \; \blockbase \; \hoare{\postcond} \; \redexstep \; B 
      } 
      \\ \\
      \inference[Synth-Loop]{
        \fresh{\sv} & \sui \in (\candVS \cap \staleVS) \\
        \staleVS \; \vdash \; \hoare{\precond} \; \concfor \{(\sv : \sui)\} \blockbase  \; \hoare{\concforeach \set{\pair{\svi}{\sui}} \phi \Invand \postcond} \; \redexstep \; B \\
      }{
        \staleVS \; \vdash \; \hoare{\precond} \; \blockbase \; \hoare{\concforeach \set{\pair{\svi}{\sui}} \phi \Invand \postcond} \; \redexstep \; B
      } \\ \\
      \inference[Assign]{
        \fresh{\svpr} \\
        \staleVS \cup \{\sv\} \; \vdash \; \hoare{ \exists \, \svpr \, . \, \precond[\sv \mapsto \svpr] \Invand \sv = E[\sv \mapsto \svpr]} \; B \; \hoare{\postcond} \; \redexstep \; B' \\
      }{
        \staleVS \; \vdash \; \hoare{\precond} \; \sv \; \Assign \; E \; \blocksep B \; \hoare{\postcond} \; \redexstep \; \sv \; \Assign \; E \; \blocksep B'  
      } \\ \\
      \inference[Put]{
        \fresh{\svpr} \\
       \staleVS \cup \{\sv\} \; \vdash \; \hoare{\exists \, \svpr \, . \, \precond[\sv \mapsto \svpr] \Invand \sv = E[\sv \mapsto \svpr]} \; B \; \hoare{\postcond} \; \redexstep \; B' \\
      }{
        \staleVS \; \vdash \; \hoare{\precond} \; \sv \; \Put \; E \; \blocksep B \; \hoare{\postcond} \; \redexstep \; \sv \; \Put \; E \; \blocksep B'  
      } \\ \\
      \inference[Inv]{
        \staleVS \; \vdash \; \hoare{\precond} \; \blockbase \; \hoare{\postcond} \; \redexstep \; B' \\
        \globalVS \; \vdash \; \hoare{\postcond} \; B  \; \hoare{\postcond} \; \redexstep \; B'' \\
      }{
        \staleVS \; \vdash \; \hoare{\precond} \; B \; \hoare{\postcond} \; \redexstep \; B' \concat  B''
      } 
      \\ \\      
      \inference[For-Extend]{
        \sui \metarel_{\precond} \su & \fresh{\sv} & \su \notin \set{\sui} \\
        \staleVS \; \vdash \; \hoare{\precond} \; \concfor \; \set{\pair{\svi}{\sui}} \cup \{\pair{\sv}{\su}\} \; B_i \; \blocksep \; B  \; \hoare{\postcond} \; \redexstep \; B' 
      }{
        \staleVS \; \vdash \; \hoare{\precond} \; \concfor \; \set{\pair{\svi}{\sui}} \; B_i \; \blocksep \; B  \; \hoare{\postcond} \; \redexstep \; B' 
      } \\ \\
      \inference[Foreach-Extend]{
        \set{\syi} \cap \set{\sui} \neq \varnothing \\ 
        \phi' = \merge{\concforeach \set{\pair{\sai}{\syi}} \; \phi_l}{\concforeach \set{\pair{\svi}{\sui}} \; \phi_r} \\
        \globalVS \; \vdash \; \hoare{\phi' \Invand \postcond} \; B  \; \hoare{\phi' \Invand \postcond} \; \redexstep \; B' 
      }{
        
          \begin{array}{@{}c@{}} 
            \globalVS \; \vdash \; \\
            \hoare{\concforeach \set{\pair{\sxi}{\syi}} \; \phi_l \Invand \concforeach \set{\pair{\svi}{\sui}} \; \phi_r \Invand \postcond} \\ 
            B \\ 
            \hoare{\concforeach \set{\pair{\sxi}{\syi}} \; \phi_l \Invand \concforeach \set{\pair{\svi}{\sui}} \; \phi_r \Invand \postcond} \\
            \redexstep \; B' 
          \end{array} 
      } \\ \\
      \inference[For-Specialize]{
        \set{\sui} \subseteq \set{\syi} & \phi' = \weakenprev{\phi} \\
        \modify{B_i} \; \vdash \; \hoare{ \phi'[\svi \mapsto \sxi] \Invand \postcond} \; \blockbase \; \hoare{ \phi[\svi \mapsto \sxi] \Invand \postcond} \; \redexstep \; B_{pre} \\
        \globalVS \; \vdash \; \hoare{\ \phi[\svi \mapsto \sxi] \Invand \postcond} \;  B_i  \; \hoare{\phi[\svi \mapsto \sxi] \Invand \postcond} \; \redexstep \; B_i' \\ 
        \globalVS \; \vdash \; \hoare{\concforeach \set{\pair{\svi}{\sui}} \; \phi \Invand \postcond} \; B  \; \hoare{\concforeach \set{\pair{\svi}{\sui}} \; \phi \Invand \postcond} \; \redexstep \; B'
      }{
        \begin{array}{@{}c@{}} 
          \globalVS \; \vdash \; \\
          \hoare{\concforeach \set{\pair{\svi}{\sui}} \; \phi \Invand \postcond } \\ 
          \concfor \; \set{\pair{\sxi}{\syi}} B_i \; \blocksep B \\ 
          \hoare{\concforeach \set{\pair{\svi}{\sui}} \; \phi \Invand \postcond} \\
          \redexstep \; \concfor \; \set{\pair{\sxi}{\syi}} (B_{pre} \concat B_i') \; \blocksep \; B'
        \end{array} 
      } \\ \\
      \inference[Conditional]{
        \globalVS \; \vdash \; \hoare{E \Invand \postcond } \; B_t \; \hoare{\postcond} \; \redexstep \; B_t' \\
        \globalVS \; \vdash \; \hoare{\lnot E \Invand \postcond } \; B_f \; \hoare{\postcond} \; \redexstep \; B_f' \\
        \globalVS \;  \; \vdash \; \hoare{\postcond } \; B \; \hoare{\postcond} \; \redexstep \; B' \\
      }{
        \globalVS \; \vdash \; \hoare{\postcond} \; \concif \; E \; \concthen \; B_t \; \concelse \; B_f \; \blocksep B \; \hoare{\postcond} \;  \redexstep \concif \; E \; \concthen \; B_t' \; \concelse \; B_f' \; \blocksep B'
      }
    \end{stackCC}
    \]
    \caption{Inference rules for \tool algorithm, with explicit blocks.}
  \label{fig:synth-rules}
\end{figure}


\paragraph{Patch Generation}
The rule {\tt Synth-Base} fires once we reach the end of the input block
and performs the actual patch generation.
It non-deterministically picks a patch satisfying the specification,
and can only update ``stale'' variables,
which are not modified
but depend on modified variables via the specification $\precond$
(we formalize this dependency in \autoref{fig:synth-mod-rel}).
Our implementation realizes the non-deterministic choice via constraint-based synthesis
in the space of all blocks that only contain assignments and put-statements. 
{\tt Synth-Loop} is similar to {\tt Synth-Base} but allows generating looping patches
when the postcondition contains quantification.


\paragraph{Accumulating Modifications} 
~
{\tt Assign} and {\tt Put} simply accumulate modifications made by the input block. In these rules, the variable modified by the current statement is added to $\staleVS$, and
the precondition of the subproblem is updated to reflect the result of the modification.
Note that while the top-level completion problem is always \emph{symmetric} 
(\ie of the form $\staleVS \; \vdash \; \hoare{\precond} \; B \; \hoare{\precond}$,
where $\precond$ is the data invariant we are trying got enforce),
the pre- and the post-condition might become different as a result of applying {\tt Assign} or {\tt Put}.
Sometimes these differences must be reconciled,
because rules like {\tt For-Specialize} only apply to symmetric goals.
The rule {\tt Inv} allow us to do just that:
restore the invariant $\postcond$ by inserting a patch in the middle of a block. 

\paragraph{Targeting} 
The central rule of our system is {\tt For-Specialize}.
If a data invariant and a loop have the same syntactic structure (i.e. iterate over the same collections), 
this rule \emph{targets} the data invariant to the loop body:
\ie strips both loop and quantification from the subgoal.
One complication here is the role of $\concprev$ terms. 
As discussed in \autoref{sec:language}, terms with $\concprev$ cannot be used as an assumption for the body of a targeted loop. 
In this case, we first patch the current loop iteration into the term $B_{pre}$, and then continue to the remainder of the loop body. 

\paragraph{Alignment} Finally, a crucial necessity for the {\tt For-Specialize} rule is that the data invariant and the loop are syntactically similar. To reach this state, the {\tt Foreach-Extend} and {\tt For-Extend} rules
syntactically search for an alignment. Both of these rules are semantics-preserving and are performed so that the Targeting rule can be applied.

\paragraph{Patching the Example}
We next give a derivation for a patch for the running example, in which we extend the loop by iterating over \T{weeks} and introduce
a maintenance Put to the new \T{weeks} iterator. 

Recall that we wish the unit-conversion invariant
on lines 4 and 5 to be a data invariant for the body of \T{adjustForCOLA}, lines 8 through 10.

In this case, the pre- and post-conditions are 
$$ \T{foreach w in weeks, d in days: 7 * d.val = w.val}, $$ 
and the block to be patched is 
$$ \T{for d in days: if (d.val > 0): d <- d.val * cola;}.$$

First, to make the loop iterate over the same variables as the \T{foreach} term,
we introduce a new iterator over \T{weeks} by applying {\tt For-Extend}, producing the new loop
\T{for w in weeks, d in days: ...}. 

Next, we target the specification to the loop by applying {\tt For-Specialize}, which has
the effect of stripping the \T{foreach} and \T{for} terms. As a consequence our new 
data invariant is \T{7 * d.val = w.val}, and our new block is \T{if (d.val > 0): d <- d.val * cola}.

We next apply {\tt Conditional} to simplify the loop. The false-branch is empty and so satisfies
the data invariant.
We now only need to patch the true-branch. 

Because the statement is a Put, we apply the {\tt Put} rule, which logically embeds
the effects of \T{d <- d.val * cola} into the precondition, and adds \T{d} to the 
set of modified variables $\staleVS$. At this point, we're left with a logical specification,
an empty block, and a set of modified variables with just one member, $\staleVS = \{d\}$.

Finally, we apply two rules. First, we find a maintenance patch for the data invariants
by the {\tt Synth-Base} rule. This produces a snippet $B'$ (in this case \T{w <- d.val * 7;}) such that if we add $B'$ at line 11,
the resulting conditional (and loop) will maintain the invariant. We will discuss this further in a moment, but for now,
we will produce an extension from $B'$ and the current block \T{d <- d.val * cola;} using 
the {\tt Inv} rule. 

Now we demonstrate how to find $B'$ using the {\tt Synth-Base} rule. 
In this case, because \T{w} and \T{d} both appear in the precondition, and \T{d} is in $\staleVS$ the candidate variables for a patch
are $\{w,d\}$. However, since $B'$ is not allowed to modify any of the variables in $\staleVS$ (i.e. \T{d}), it's
forced to produce a patch that modifies \T{w}, which further satisfies the invariant
\T{w.val = d.val * 7}. One such patch is \T{w <- d.val * 7;}, and so the {\tt Synth-Base} rule
calculates this patch for $B'$.

\begin{figure}[t]
  \centering
  \scriptsize
  \[
  \begin{stackCC}
    \inference[Rel-Expr]{
      \atomic{\pi} & \sx, \sy \in \free{\pi} \\
    }{
      \sx \metarel_{\pi} \sy
    } \quad
    \inference[Rel-Left]{
      \sx \metarel_{L} \sy   \\
    }{
      \sx \metarel_{L \Invand R} \sy
    }\\ \\
    \inference[Rel-Right]{
      \sx \metarel_{R} \sy   \\
    }{
      \sx \metarel_{L \Invand R} \sy
    } \quad
    \inference[Rel-Trans]{
      \sx \metarel_{L} \sy & \sy \metarel_R \st  \\
    }{
      \sx \metarel_{L \Invand R} \st
    }
  \end{stackCC} 
  \]
  \caption{Inference rules for variable data-dependency relation. We relate two variables $x$ and $y$ by $\metarel$ if a modification to $x$ might affect $y$.}
  \label{fig:synth-mod-rel}
\end{figure}

\subsection{Soundness of Synthesis Rules}\label{subsec:synth-soundness}
In all cases, if the \tool extension rules produce a new program, the program must satisfy the input data invariants. 
We formalize the synthesis soundness using the axiomatic semantics of \autoref{sec:language}: 

\begin{theorem}[Soundness of \technique]

  \[ \forall \, \precond, \, B, \, B' \, .\, \emptyset \, \vdash \hoare{\precond} \, B \, \hoare{\precond} \, \redexstep \, B' \quad \implies \hoare{\precond} \, B' \, \hoare{\precond}
  \]

\end{theorem}

We prove this by generalizing to $\staleVS \, \vdash \hoare{\precond} \, B \, \hoare{\postcond} \, \redexstep \, B'$
and then by induction on the derivation. 
More detail is in \autoref{app:proof-synth-sound} and the proof is straightforward.

%% file: evaluation.tex
\section{Evaluation}\label{sec:evaluation}

In this section, we detail the experiments run to evaluate \tool.
We assessed \tool quantitatively via a set of benchmarks and using several case studies.

\paragraph{Research questions}
We test the following questions:

\begin{compactenum}[~(RQ1)]
\item Is programming with \tool and data invariant is more succinct (and therefore easier) than maintaining data invariants manually?\label{rq1}
\item Does \tool make code evolution easier? We test this by examining the necessary changes to implementation and invariants in order to implement new functionality.\label{rq2}
\item Does \technique enable fast, scalable synthesis? To test this, we measure the performance of synthesizing with \tool.\label{rq3}
\end{compactenum}

\paragraph{Implementation}
We evaluate \tool and \technique using a prototype compiler that targets Boogie \cite{leino2008boogie}. 
Our prototype implements the contents of \autoref{sec:language} by compiling to Boogie,
and we implement the contents of \autoref{sec:synthesis} by extending \tool terms using our own synthesis and CEGIS algorithms.

\subsection{Case Studies}
We first examine RQ\ref{rq1} and RQ\ref{rq2} using three detailed case studies.

The invariant language of \tool, targeted towards expressing relations over collections,
is a perfect fit for many useful idioms in web programming. 
Using \tool, we implemented three applications inspired by real-life web programs.

\subsubsection{Game of Life}
John Conway's Game of Life \cite{conway1970game} 
is a popular visualization of a cellular automaton with applications in Chemistry, Physics, Math, and Computer Science.
In this game, a discrete world of cells obeys particular evolutionary behavior. 
At each time step of the application, the cells in the 
world change state according to the rules of the game. 
We looked at several interactive applications of the game of life online, such as \cite{GoLOnline}. 
In all of these applications, the programmer manually maintained an invariant between
the visual cells of the board and the internal data structure for the cells.
To implement this in \tool, we encoded the internal state of the game and its visual state as two integer arrays.
An element-wise invariant relates the internal state of the game to its visual state.
We implemented procedures for 
\begin{inparaenum}
\item making transitions in the internal state according to the rules of the game,
\item interactive logic that allows the user to change the state of a cell by clicking on the board, and 
\item a buttom for starting and stopping the game.
\end{inparaenum}
\tool was able to synthesize a patch that re-synchronizes the model and the view 
for each of these procedures.

\subsubsection{Budgeting Application}
Our second case study is a spreadsheet-style budgeting application, 
described in detail in \autoref{sec:motivating-example}.
For this benchmark, the programmer builds a financial application which takes in periodic revenues and deficits.
This application takes amounts in three periodic intervals---weekly, monthly, and yearly---and converts between the
amounts. In this way, the end-user can input data in the most convenient format. 

A difficult feature of this benchmark was summing up the rows of the budget and presenting a total value.
In traditional programming, this would require a procedure and would not be easy to compose.
In contrast, in \tool, this invariant is easily expressible using the \T{prev} calculus and indeed composes very well with the other invariants of the system.

\subsubsection{Shared Expenses Application}
Our final case study is an extension of the Budgeting Application. 
Anecdotally, one of the co-authors actually uses this type of application in real-life.
The idea here is that two people who live in the same household want to split shared expenses equally at the end of the month. 
In this application, each row has 4 entries: in the first two cells store the expenses paid by person A and person B, respectively;
in the third cell, stores the average cost for the expense 
(i.e. the final cost for each person), 
and in the fourth cell, the amount person A owes to person B 
(i.e. how much person B over/underpaid on the particular expense).
Similar to the budgeting application, we can express each row of this application in \tool and further, we can 
conditionally render the amount owed between the participants. 

\subsection{Quantitative Evaluation}

In addition to the qualitative evaluation, we empirically evaluate questions $1$-$3$ on a series of benchmarks and compare them to two traditional techniques,
manual invariant maintenance, and dynamic maintenance of functional specifications (i.e. Functional-Reactive Programming, FRP).

To compare \tool against these two techniques in a language-agnostic, apples-to-apples way, we 
implement each benchmark in all three paradigms using \tool's syntax. For the imperative
paradigm, we manually maintain invariants without using specifications. For the FRP paradigm,
we write functional specifications for each variable in the program. 

\input{results.tex}

\paragraph{RQ\ref{rq1}: Succinctness}
We measure the amount of code necessary to implement a set of benchmarks in three different techniques: manually (Imperative), FRP and \tool.
We show the results in \autoref{fig:evaluation-table}.
As expected, the size of manually implemented code for both FRP and \tool is considerably smaller than Imparative. 
However, \tool specifications are as much as three times smaller than FRP specifications.
Additionally, we see that patches are generated in a number of locations.
This means manually maintainaing the invariants would have required to keep track of all these locations.
We do see that the size of the patches generated by \tool is much larger than the size of the manual implementation. There are two main reasions fo this:
\begin{inparaenum}[1)]
\item \tool patches are not meant for human consumption and so are unoptimized, and
\item patches are synthesized in the target language (i.e., Boogie), which is not as concise as \tool.
\end{inparaenum}
The results show that \tool invariants provide a succinct way of specifying what would otherwise be a much larger piece of enforcement code.

\begin{figure}[t]
      \centering

      \begin{subfigure}[t]{0.45\columnwidth}
            \centering
            \includegraphics[trim=4 4 4 4,clip,width=\textwidth]{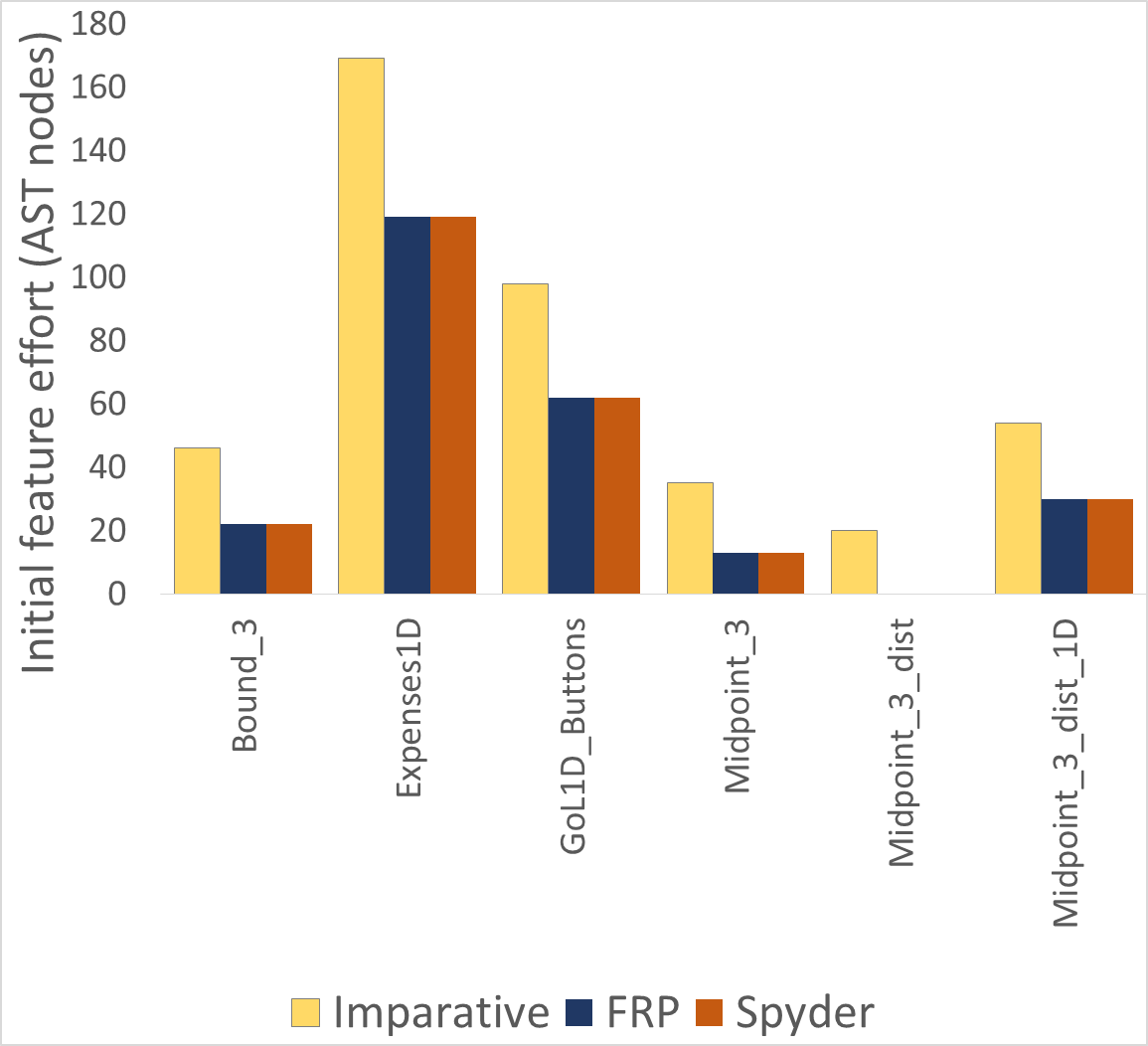}
            \caption{Initial implementation effort: diff size to implement a new feature (in AST nodes) while breaking invariants in all three techniques.}\label{fig:graphs-baseline}
      \end{subfigure}%
      ~
      \begin{subfigure}[t]{0.45\columnwidth}
            \centering
            \includegraphics[trim=4 4 4 4,clip,width=\textwidth]{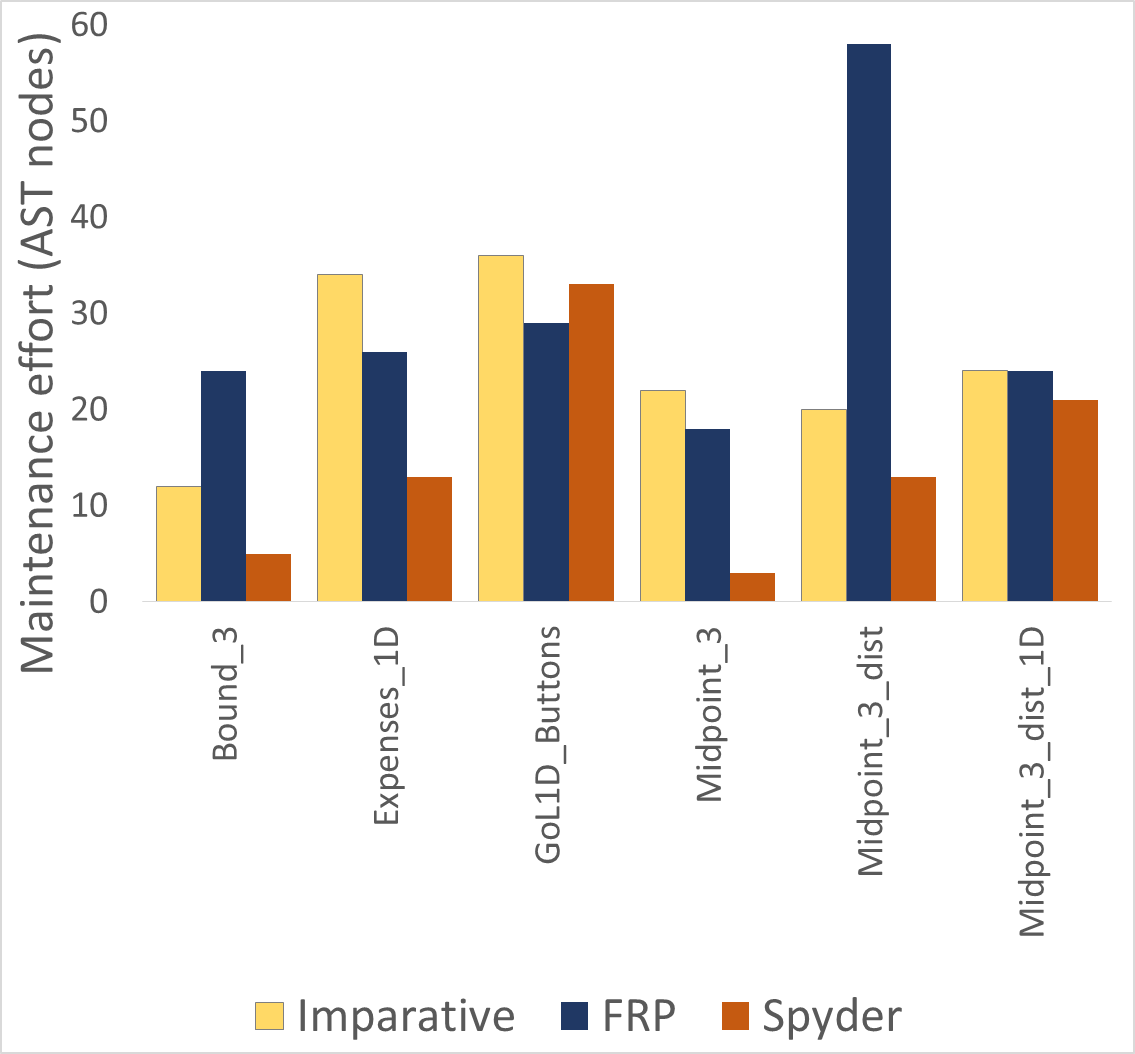}
            \caption{Invariant maintenance effort: effort repairing broken invariants. For FRP and \tool indicates updates to invariants and for Imperative this is manually written invariant maintenance code.}\label{fig:graphs-maintenance}
      \end{subfigure}

      \caption{
            Effort required to add new features to existing programs. 
      }
      \label{fig:evaluation-graphs}
\end{figure}

\paragraph{RQ\ref{rq2}: Ease of modification} 
We measure the amount of modification required to evolve existing code, again comparing \tool to imperative and FRP invariant maintenance.
\autoref{fig:graphs-baseline} shows for each benchmark the size of the modification (in AST nodes) required to implement the new functionality portion of a new feature, without fixing broken data invariants, and \autoref{fig:graphs-maintenance} shows the size of the modification to invariant-preserving code.

As seen in our discussion of RQ\ref{rq1}, we see in \autoref{fig:graphs-baseline} that writing new functionality with \tool is more succinct than either writing it imperatively in the \tool language.
\autoref{fig:graphs-maintenance} shows that the same is true for invariant maintenance: modifications to the invariants in \tool are considerably smaller than the manual changes in imp and the changes to code and invariants in FRP.

These benchmarks show that code evolution is also easier in \tool.

\paragraph{RQ\ref{rq3}: Performance} We evaluated the scalability of the \tool compiler (and by extension, the \technique algorithm) by compiling our benchmarks and comparing the performance against 
a standard synthesis technique, Sketch \cite{solar2008program}. For each of our benchmarks, we reimplemnted the benchmark in Sketch and compared the performance. 
In contrast to \technique, Sketch performs bounded enumeration for verification, and as a consequence, quantified invariants scale in proportion to the size of the verified array.
To measure the scalability of bounded verification, for Sketch programs with arrays we varied the number of elements in the concrete Sketch arrays from 3 elements to 50 elements. 

Overall, we find that Sketch outperforms \tool on the synthetic benchmarks but does not complete within the time limit in two of of our three case studies. For the case studies,
Sketch could solve these problems if the programmer wrote a synthesis template tailored to the specific study. In contrast, \tool programmers do not have to develop
an application-specific sketch.

%% file: results.tex
\newcommand{\head}[1]{\emph{#1}}
\newcommand{\timeout}{XXX\xspace}

\begin{table}[t]
\centering
\scriptsize
\begin{tabular}{@{} c|l|cc|cc|cc|ccc|c|cc| @{}}
 & \multirow{2}{*}{Benchmark} & \multicolumn{2}{c|}{Imp Size} & \multicolumn{2}{c|}{FRP Size} & \multicolumn{2}{c|}{Spy Size} & \multicolumn{4}{c|}{Synthesis Time} & \multicolumn{2}{c|}{Patches}\\
                            \cline{3-14} 
      &      & \head{Impl} & \head{Spec} & \head{Impl} & \head{Spec} & \head{Impl} & \head{Spec} & \head{SK-3} & \head{SK-10} & \head{SK-50} & \head{Spy} & locs & size\\	

\hhline{==============}

\multirow{6}{*}{\rotatebox[origin=c]{90}{\tiny Arithmetic}}  & Midpoint\_2 & 39 & 0 & 27 & 13 & 27 & 7 & 1.04 & N/A & N/A & 58.01 & 2 & 298 \\
& Midpoint\_3 & 64 & 0 & 40 & 25 & 40 & 9 & 1.44 & N/A & N/A & 27.78 & 3 & 123 \\
& Midpoint\_3\_dist & 76 & 0 & 40 & 81 & 40 & 22 & 0.80 & N/A & N/A & 123.24 & 3 & 489 \\
& Midpoint\_3\_dist\_1D & 127 & 0 & 67 & 105 & 67 & 42 & 32.52 & 322.82 & 303.26 & 97.79 & 3 & 483 \\
& Bound\_2 & 39 & 0 & 27 & 13 & 27 & 5 & 0.36 & N/A & N/A & 116.20 & 2 & 198 \\
& Bound\_3 & 64 & 0 & 40 & 37 & 44 & 10 & 0.92 & N/A & N/A & 348.68 & 3 & 418 \\

\hline\multirow{4}{*}{\rotatebox[origin=c]{90}{\tiny Web Apps}} & GoL1D & 279 & 0 & 255 & 25 & 255 & 13 & 0.74 & 0.84 & 1.15 & 38.77 & 4 & 30 \\
& GoL1D\_Buttons & 373 & 0 & 317 & 54 & 317 & 46 & 1.08 & 0.27 & 1.29 & 49.38 & 8 & 194 \\
& Expenses & 59 & 0 & 43 & 17 & 43 & 9 & 29.09 & N/A & N/A & 107.58 & 2 & 233 \\
& Expenses\_1D & 170 & 0 & 126 & 37 & 126 & 19 & - & - & - & 105.93 & 3 & 190 \\
& Overview & 105 & 0 & 63 & 151 & 63 & 82 & - & - & - & 106.25 & 2 & 820 \\
 \hhline{==============}
\end{tabular}

\caption{
Benchmarks comparing implementation in \tool to other techniques.
\head{Impl} is the size of implementation (non-invariant) code and \head{Spec} is the size of invariants. 
All sizes are in AST nodes, and all implementations are in the \tool language.
Each benchmark has invariants maintained manually (\head{Imp}), in the \head{FRP} paradigm, and with \tool.
Synthesis times of \tool are given compared to Sketch (\head{SK}), on collections of size \head{3}, \head{10}, and \head{50} (\head{SK-3}, \head{SK-10} and \head{SK-50} resp.).
We report a timeout ($-$) after ten minutes and we use N/A to denote a Sketch program that doesn't use collections.
\head{Patches} reports the size of patches synthesized by \tool and the number of patches (\head{locs}) per benchmark.
}
\label{fig:evaluation-table}
\end{table}

%% file: related.tex

\section{Related Work}\label{sec:related}

This paper builds upon two lines of prior work, 
which until now have developed independently:
\emph{declarative constraint programming}, 
where the goal is to enforce global constraints at run time,
and \emph{program synthesis} and \emph{repair}, 
which enforces traditionally local, end-to-end functional specifications at compile time.
We first discuss the trade-offs between static and dynamic constraint solving, and then we detail each of these areas.

\para{Static and Dynamic Constraint Solving}
Two of the longstanding research problems for constraint solving are performance \cite{freeman1990incremental,Badros2001,felgentreff2015checks}, as well as debugging over- and under-constrained systems \cite{felgentreff2015checks,Hottelier2014,schulte1996oz,meier1995debugging}.
In essence, the choice of static vs dynamic constraint solving boils down to a tradeoff between issues at compile
time vs issues at run time.

For performance, solving constraint statically results in (notoriously) long synthesis and compilation times, but produces fast code. 
Conversely, dynamic constraint solving does not require an expensive compilation pass but results in large runtime overheads, as high as 10x-100x (as reported in \cite{felgentreff2015checks}). 
Consequently, the choice of static vs. dynamic for performance is a tradeoff between compilation time and runtime performance.

Debugging constraint systems is a similar story in that static systems can report a compile-time error when the system is over- or under-constrained. 
Conversely, dynamic systems generally attempt to resolve ill-posed systems anyway, using techniques such as constraint hierarchies \cite{borning1992constraint}, which results in unintuitive solutions -- unintuitive because the solution does not satisfy the constraints. 
In either case, the ill-posed system must be debugged. 
In the static case, it is strictly the programmer who debugs the system, while in the dynamic case, the end user might be exposed to the ill-posed system. 
Consequently, the choice of static vs. dynamic for debugging is a tradeoff between programmer time and user time.

\para{Dynamic Invariant Enforcement}
There are two closely related research arcs on dynamically enforcing invariants:
the field of constraint imperative programming, and the work of functional reactive programming.
Both of these areas provide mechanisms for dynamically solving invariants, and both are orthogonal
to our efforts because we solve constraints statically through program synthesis. 
\paragraph{Constraint Imperative Programming.}
The field of constraint solving is rich and storied \cite{badros2001cassowary,dantzig1955generalized}, as 
constraint solvers excel at calculating global solutions. Despite their power, constraint solvers are traditionally relegated to libraries. 
The field of Constraint Imperative Programming aims to provide first-class
language support for constraint solving \cite{grabmuller2004turtle,felgentreff2015checks,oney2012constraintjs}, but again, fundamentally our work is orthogonal because
we solve constraints statically.
\paragraph{Functional Reactive Programming}
The field of Functional Reactive Programming (FRP) provide a dataflow language for building graphical systems \cite{wan2000functional}. 
Although inspired by animations, FRP quickly became popular as a tool for taming web application logic \cite{meyerovich2009flapjax,cooper2006embedding}.
The most popular recent work in this field are Elm \cite{czaplicki2013asynchronous} and its imperative cousin React \cite{staff2016react}, 
which provide a language and runtime for building client-side web applications. Although popular and powerful, FRP is a general,
dynamic technique for abstracting over dataflow -- in contrast, our work focuses on the problem of first-class data invariants, and solves
for invariant patches statically.

\para{Program Synthesis and Repair}
In recent years, \emph{program synthesis} has emerged as a promising
technique for automating tedious and error-prone aspects of programming~\cite{GulwaniHS12,Solar-Lezama:STTT13,TorlakB14}.
The two main directions in this area are
synthesis from informal descriptions (such as examples, natural language, or hints)%
\cite{Albarghouthi-al:CAV13,Polozov-Gulwani:OOPSLA15,Osera-Zdancewic:PLDI15,Feser-al:PLDI15,Smith-Albarghouthi:PLDI16,Feng-al:PLDI17,Yaghmazadeh-al:OOPSLA17,Murali-al:ICLR18,chugh2016programmatic}
and synthesis from formal specifications,
where the goal is to synthesize a program that is provably correct relative to the specification%
~\cite{Gulwani-al:PLDI11,Srivastava-al:POPL10,Kneuss-al:OOPSLA13,Delaware-al:POPL15,Polikarpova-al:PLDI16,leino2012program}.
Both directions have been mainly focusing on synthesizing standalone programs 
from complete end-to-end functional specifications of their inputs and outputs.
Instead, the present work focuses on synthesizing code snippets from incomplete, global specifications (data invariants)
and integrating them with hand-written code.

The only prior techniques we are aware of for generating snippets from declarative specifications
and inserting them into hand-written code is the in the context of information-flow security%
~\cite{FredriksonJJRPSY12,PolikarpovaYIS16}.
Enforcement of data invariants brings a different set of challenge,
since invariants are deep semantic properties.

Our work is related to sound program repair~\cite{KneussKK15},
where the problem is, given a formal specification and a program that violates it,
modify the program so that it provably satisfies the specification.
Program repair, however, is a very general problem,
and so lacks a-priori restrictions on modifications the algorithm is allowed to make.
As a result, if the given specification is incomplete, the problem is ill-defined.
In this work we show that in the setting of enforcing data invariants,
the space of possible modifications can be sufficiently restricted
to make repair both \emph{predictable} and \emph{efficient}.
Where efficiency is concerned, the deductive program repair technique of~\cite{KneussKK15}
does not scale with the number of patches generated in one function,
whereas \tool leverages the restrictions 
to solve each synthesis task independently,
hence avoiding a combinatorial explosion with the number of patches.

\para{Program Verification}
The programming and invariant language of \tool is purposefully simple,
allowing us to explore the idea of automatic invariant maintenance
without getting distracted by challenges of program verification 
in the presence of aliasing, dynamic object structures, and arbitrary quantified invariants.
There is a rich body of prior work in program verification that deals with these challenges,
both in general~\cite{Reynolds02,DynFrames} and in the specific context of object invariants~\cite{Barnett04,Dynamic,Friends,History,Muller06,Middelkoop08,Considerate,PolikarpovaTFM14}.
Extending \technique to support one of these verification methodologies is an interesting direction for future work,
but we consider it orthogonal to the initial exploration of programming with invariants.

%% file: appendix.tex
\appendix
\section{Appendix}

\subsection{ImpArray Syntax and Semantics}\label{app:imp-semantics}

\begin{figure}[h]
  \centering
    \[
      \begin{array}{lrl}
        &         & \boogv, \boogu \in \boogVars, \quad \boogi \in \mathbb{Z} \\
                   \boogStmt  & \bnfdef & \boogv \; \Assign \; \boogExpr \\
                & \bnfalt & \boogv \; [ \; \boogExpr \; ] \; \Assign \; \boogExpr \\
                & \bnfalt & \cif \; \boogExpr \; \cthen \; \boogStmt \; \celse \; \boogStmt \\ 
                & \bnfalt & \concwhile \; \; \boogExpr \; \boogStmt \\ 
                & \bnfalt & \boogStmt \; \blocksep \; \boogStmt \\ 
                & \bnfalt & \blockbase \\
        \boogExpr  & \bnfdef & \boogv \bnfalt \boogi \bnfalt \conctrue \bnfalt \concfalse \\
                & \bnfalt & \boogExpr \; \boogbop \; \boogExpr \\
                & \bnfalt & \cif \; \boogExpr \; \cthen \; \boogExpr \; \celse \; \boogExpr \\
                & \bnfalt & \booguop \; \boogExpr\\
                & \bnfalt & \boogExpr \; [ \; \boogExpr \; ] \\
                & \bnfalt & \concsize (\boogv)\\
                & \bnfalt & \forall \; \boogv \; . \;  \boogExpr \\
                & \bnfalt & \exists \; \boogv \; . \;  \boogExpr \\
        \boogbop   & \bnfdef & \plus \bnfalt \times \bnfalt \% \bnfalt \implies \bnfalt \Longleftrightarrow \bnfalt \ldots \\
        \booguop   & \bnfdef & \lnot \bnfalt ! \\
      \end{array}
    \]
    \caption{Syntax for the \targetLang language.}
    \label{subfig:imp-syntax}

\end{figure}

\begin{figure}[h]
  \[
  \begin{array}{lcc}
    \llbracket \boogv \rrbracket_{\sigma}     & = & \sigma[\boogv] \\
    \llbracket \boogi \rrbracket_{\sigma}     & = & \boogi \\
    \llbracket \conctrue \rrbracket_{\sigma}  & = & \top \\
    \llbracket \concfalse \rrbracket_{\sigma}  & = & \bot \\
    \llbracket E_l \, \boogbop \, E_r  \rrbracket_{\sigma}  & = & \llbracket E_l \rrbracket_{\sigma} \boogbop \llbracket E_r \rrbracket_{\sigma} \\
    \llbracket \booguop \, E_i  \rrbracket_{\sigma}  & = & \booguop \llbracket E_i \rrbracket_{\sigma} \\
    \llbracket \boogv [E] \rrbracket_{\sigma}  & = &  \llbracket \boogv \rrbracket_{\sigma} [\llbracket E \rrbracket_{\sigma}] \\
    \llbracket \concsize(E) \rrbracket_{\sigma}  & = &  \| \llbracket E \rrbracket_{\sigma} \| \\
    \llbracket \forall \boogv \, . \, E \rrbracket_{\sigma}  & = &  \forall \, x \in \sigma \, . \, \llbracket E[v \mapsto x] \rrbracket_{\sigma} = \top \\

  \end{array}
  \]

  \caption{Denotational semantics for \targetLang expressions}
  \label{subfig:imp-semantics-expr}

\end{figure}

\begin{figure}[h]
    \[
    \begin{stackCC}
      \inference {
      } {
        \boogv \; \Assign \; \boogExpr , \; \sigma \; \rightsquigarrow \sigma[\boogv \mapsto \llbracket \boogExpr \rrbracket_{\sigma}]
      } \\ \vspace{3mm}
      \inference {
      } {
        \boogv \; [ \; \boogExpr_1 \; ] \; \Assign \; \boogExpr_2 \; , \; \sigma \; \rightsquigarrow \sigma[\boogv \mapsto \boogv[\llbracket \boogExpr_1 \rrbracket_{\sigma} \mapsto \llbracket \boogExpr_2 \rrbracket_{\sigma}]
      } \\ \vspace{3mm}
      \inference {
        \llbracket \boogExpr \rrbracket_{\sigma} \; = \; \top \\
        \boogStmt_1 \; , \; \sigma \rightsquigarrow \sigma'
      } {
        \cif \; \boogExpr \; \cthen \; \boogStmt_1 \; \celse \; \boogStmt_2 \; , \; \sigma \; \rightsquigarrow \sigma'
      } \\ \vspace{3mm}
      \inference {
        \llbracket \boogExpr \rrbracket_{\sigma} \; = \; \bot \\
        \boogStmt_2 \; , \; \sigma \rightsquigarrow \sigma'
      } {
        \cif \; \boogExpr \; \cthen \; \boogStmt_1 \; \celse \; \boogStmt_2 \; , \; \sigma \; \rightsquigarrow \sigma'
      } \\ \vspace{3mm}
      \inference {
        \llbracket \boogExpr \rrbracket_{\sigma} \; = \; \top 
      } {
        \concwhile \; \; \boogExpr \; \boogStmt \; , \; \sigma \; \rightsquigarrow \sigma
      } \\ \vspace{3mm}
      \inference {
        \llbracket \boogExpr \rrbracket_{\sigma} \; = \; \bot \\
        \boogStmt \; \blocksep \; \concwhile \; \boogExpr \; \boogStmt \; , \sigma \; \rightsquigarrow \sigma' 
      } {
        \concwhile \; \; \boogExpr \; \boogStmt \; , \; \sigma \; \rightsquigarrow \sigma
      } \\ \vspace{3mm}
      \inference {
        \boogStmt_1 \; , \; \sigma \; \rightsquigarrow \; \sigma'
        \boogStmt_2 \; , \; \sigma' \; \rightsquigarrow \; \sigma''
      } {
        \boogStmt_1 \; \blocksep \; \boogStmt_2 \; , \; \sigma \; \rightsquigarrow \sigma''
      } \\ \vspace{3mm}
      \inference {
      } {
        \blockbase \; , \; \sigma \; \rightsquigarrow \sigma
      }
    \end{stackCC}
    \]
  
    \caption{Operational semantics for \targetLang statements}
    \label{subfig:imp-semantics-stmt}
  
\end{figure}

\subsection{Spyder Semantics}\label{app:spy-semantics}

Let $\sigma$ be a \targetLang state, $E$ a \tool Expression, and $\Gamma$ a well-formed translation context with respect to $E$. We define the denotational semantics
of $E$ as the denotational semantics of the corresponding \targetLang expression:

\begin{definition}[Spyder Expression Semantics]\label{def:expr-semantics}

  \[
    \llbracket E \rrbracket_{\sigma} \quad \bnfdef \quad \llbracket \translate{E}{\Gamma} \rrbracket_{\sigma}
  \]

\end{definition}

We similarly define the operational semantics of a \tool statement $S$ as the operational semantics of the corresponding \targetLang statement $\translate{S}{\Gamma}$:

\begin{definition}[Spyder Statement Semantics]\label{def:stmt-semantics}

  \[
    \inference {
      \translate{S}{\Gamma} \; , \; \sigma \; \rightsquigarrow \; \sigma'
    } {
      S \; , \; \sigma \; \rightsquigarrow \; \sigma'
    }
  \]
\end{definition}

\begin{figure}[h]
  \centering
    \[
  \begin{stackCC}
    \inference[Consequence]{
          P \implies P' \qquad Q' \implies Q \\
          \hoarebr{P'} \; S \; \hoarebr{Q'}
        }{
          \hoarebr{P} \; S \; \hoarebr{Q}
        }  \\ \\
        \inference[Skip]{
        }{
          \hoarebr{P} \; \blockbase \; \hoarebr{P} 
        } \\ \\
        \inference[Sequence]{
          \hoarebr{P} \; S_1 \; \hoarebr{Q} \\
          \hoarebr{Q} \; S_2 \; \hoarebr{R} \\
        }{
          \hoarebr{P} \; S_1 \; \blocksep \; S_2 \; \hoarebr{R} 
        } \\ \\
        \inference[Conditional]{
          \hoarebr{P \land e} \;  S_t \; \hoarebr{Q} \\
          \hoarebr{P \land \lnot \; e } \; S_f \; \hoarebr{Q} \\ 
        }{
          \hoarebr{P} \; \Cond{e}{S_t}{S_f} \; \hoarebr{Q}
        } \\ \\
        \inference[Assign-Var]{
          \fresh{v'} \\
        }{
          \hoarebr{P} \; v  \; \Assign \; E \; \hoarebr{ \, \exists v' \, . \, P[v \mapsto v'] \land v = E[v \mapsto v']} 
        }  \\ \\
        \inference[Assign-Array]{
          \fresh{v'} \\
        }{
          \hoarebr{P} \; v [ \; E_i \; ] \; \Assign \; E_r \; \hoarebr{\exists \, v' \, . \, P[v \mapsto v'] \land v = v'[E_i[v \mapsto v'] \, \Assign \, E_r[v \mapsto v']]}
        }  \\ \\
        \inference[While]{
        \hoarebr{ I \land E } \; S \; \hoarebr{ I } \\
      }{
        \hoarebr{ I } \; \concwhile \; E \; S \; \hoarebr{ I \land \lnot E}
      } \\ \\

  \end{stackCC}
  \]
  \caption{Standard axiomatic semantics (Hoare logic) for \targetLang.}
  \label{fig:imp-verification}
\end{figure}

\subsection{Soundness of Spyder Triples}\label{app:proof-axiomatic-sound}

\begin{lemma}[Bindings]\label{lem:bind}

  \[ \forall \, P, \, B, \, \Gamma \, . \wellform{\concfor \pair{\sx}{\sy} B}{\Gamma} \land \wellform{P}{\Gamma} \, \implies \, \sx \notin \free{P}
  \]

\begin{proof}
  Induction over the derivation of $\wellform{P}{\Gamma}.$
\end{proof}
\end{lemma}

\begin{lemma}[Assignment]\label{lem:assign}

  \[ \forall \, B, \, \Gamma \, . \wellform{\concfor \pair{\sx}{\sy} B}{\Gamma} \, \implies \, x \notin \Assigned{B}
  \]

\begin{proof}
  Induction over the derivation of $\wellform{\concfor \pair{\sx}{\sy}B}{\Gamma}.$
\end{proof}
\end{lemma}

\begin{lemma}[Array substitution]\label{lem:arr-sub}

  \begin{multline*} \forall \, P, \, \sx, \, \sy, \, \Gamma \, . \wellform{P}{\Gamma} \land \Gamma(\sx) = \sy \, \implies \\
   \forall \, \sigma \, . \, \sigma(\translate{P}{\Gamma}[\boogy \mapsto \boogypr]) \implies \sigma(\translate{P}{\Gamma}[\boogy[\boogx] \mapsto \boogxpr])
  \end{multline*}

\begin{proof}
  Structural induction over $P$. 
\end{proof}
\end{lemma}


\begin{theorem}[Relative Soundness] 
  \begin{multline*}
    \forall P, \, S, \, Q, \, \Gamma \, . \, \wellform{P \land Q}{\Gamma} \land \wellform{S}{\Gamma}  \implies \\
      \hoare{P} \, S \, \hoare{Q} \, \implies \hoarebr{\translate{P}{\Gamma}} \, \translate{S}{\Gamma} \, \hoarebr{\translate{Q}{\Gamma}}
  \end{multline*}

\begin{proof} By induction over the derivation of $\hoare{P} \, S \, \hoare{Q}$; for each case of $S$, we build a corresponding
derivation for $\hoarebr{\translate{P}{\Gamma}} \, \translate{S}{\Gamma} \, \hoarebr{\translate{Q}{\Gamma}}$. 


In all cases we start by assuming $\wellform{P \land Q}{\Gamma} \land \wellform{S}{\Gamma}.$

Cases of S: \begin{enumerate}
  \item Base case, in which the last step of the derivation is {\tt Skip}: $\hoare{P} \, \blockbase \, \hoare{Q}$. From the structure of {\tt Skip}, it must be the case that
  $P$ and $Q$ are structurally identical, i.e. the derivation is $\hoare{P} \, \blockbase \, \hoare{P}$. Since {\tt trans} is a function, it maps $\blockbase$ to exactly one statement (namely $\blockbase$), and $P$ to exactly one expression $\translate{P}{\Gamma}$.
  Finally, we apply the {\tt Skip} Hoare rule to obtain $\hoarebr{\translate{P}{\Gamma}} \, \blockbase \, \hoarebr{\translate{P}{\Gamma}}$.

  \item Inductive case, in which the last step of the derivation is {\tt Consequence}: $\hoare{P} \, S \, \hoare{Q}$. We will use the corresponding {\tt Consequence} rule of Hoare logic to build a derivation for $\hoarebr{\translate{P}{\Gamma}} \, \translate{S}{\Gamma} \, \hoarebr{\translate{Q}{\Gamma}}$.
  
  Since the case is {\tt Consequence}, there must be $P'$ and $Q'$ such that $P \implies P'$, $Q' \implies Q$, and $\hoare{P'} \, S \, \hoare{Q'}$.
  From \autoref{def:semantics}, we know that $\translate{P}{\Gamma} \implies \translate{P'}{\Gamma}$ and $\translate{Q'}{\Gamma} \implies \translate{Q}{\Gamma}$.
  From the inductive hypothesis, we have the ImpArray triple

  $$\hoarebr{\translate{P'}{\Gamma}} \, \translate{S}{\Gamma} \, \hoarebr{\translate{Q'}{\Gamma}},$$

  and so we apply the {\tt Consequence} ImpArray rule to obtain 

  $$\hoarebr{\translate{P}{\Gamma}} \, \translate{S}{\Gamma} \, \hoarebr{\translate{Q}{\Gamma}}.$$
  

  \item Inductive case, in which the last step of the derivation is {\tt Conditional}: $\hoare{P} \; \Cond{E}{S_t}{S_f} \; \hoare{Q}.$ This follows from the inductive hypothesis applied to $E$, $S_t$, and $S_f$, as well as the {\tt Conditional} ImpArray rule.
  \item Inductive case, in which the last step of the derivation is {\tt Sequence}: $\hoare{P} \; S_1 \; \blocksep \; S_2 \; \hoare{R}.$ This follows from the inductive hypothesis applied to $S_1$, and $S_2$, as well as the {\tt Sequence} ImpArray rule.
  \item Inductive case, in which the last step of the derivation is {\tt Assign}: $\hoare{P} \; \sv \; \Assign \; E \; \hoare{Q}.$ In this case, the translation produces
  a Imp assignment to $\sv$.
  
  Since the last step is {\tt Assign}, there must be a fresh variable $\svpr$ such $Q$ is the strongest postcondition of the assignment to $\sv$: 
  $$\exists \, \svpr \, . \, P[\sv \mapsto \svpr] \land \sv = E[\sv \mapsto \svpr]$$

  From the inductive hypothesis, we know that translating the Spyder triple produces an equivalent
  Imp Hoare triple 
  $$\hoarebr{\translate{P}{\Gamma}} \; \boogv \, \Assign \, \translate{E}{\Gamma} \, \hoarebr{\translate{\exists \, \svpr \, . \, P[\sv \mapsto \svpr] \land \sv = E[\sv \mapsto \svpr]}{\Gamma}}.$$

  If you consider the translated term $\translate{\exists \, \svpr \, . \, P[\sv \mapsto \svpr] \land \sv = E[\sv \mapsto \svpr]}{\Gamma}$, using \autoref{lem:subst} and the definition of translation,
  you'll find that it is exactly the ImpArray postcondition for assignment with $\translate{P}{\Gamma}$ as a precondition:
  $$ \exists \, \boogvpr \, . \, \translate{P}{\Gamma}[\boogv \mapsto \boogvpr] \land \boogv = \translate{E}{\Gamma}[\boogv \mapsto \boogvpr]. $$

  So, we apply {\tt Assign} with $P$ as a precondition to obtain
  \begin{multline*}
  \hoarebr{\translate{P}{\Gamma}} \boogv \, \Assign \\
  \translate{E}{\Gamma} \, \hoarebr{ \exists \, \boogvpr \, . \, \translate{P}{\Gamma}[\boogv \mapsto \boogvpr] \land \boogv \, = \, \translate{E}{\Gamma}[\boogv \mapsto \boogvpr]},
  \end{multline*}

  \item Inductive case, in which the last step of the derivation is {\tt Put}: $\hoare{P} \; \sv \Put E \; \hoare{Q}.$ For this, we will show that the translation of 
  the put $\sv \Put E$ takes the precondition $\translate{P}{\Gamma}$ to the translation of the Spyder post-condition $\exists \, \svpr \, . \, \weakenforeach{P}{v}{\Gamma}[\concelem(\sv) \mapsto \svpr] \land \concelem(\sv) = E[\concelem(\sv) \mapsto \svpr].$

  Consider the translation of $\concelem(\sv)$ in the context of $\Gamma$. Since $\Gamma$ is well-formed with respect to the Put to $\sv$, it must be the case
  that $\sv \in \Gamma$ and $\Gamma(\sv) = \boogy$ for some variable $\boogy$. Furthermore, the Spyder expressions $\concelem(\sv)$ and $\conciter(\sv)$ are translated to $\boogy[\boogv]$ and $\boogv$ respectively.

  Next, consider the Hoare postcondition of the translated put statement. The $Put$ statement is translated to $\boogy[\boogv] \, \Assign \, \translate{E}{\Gamma}$, and we can apply the {\tt Assign-Array} rule to obtain the postcondition of $\translate{P}{\Gamma}:$

  \begin{multline*}\hoarebr{\translate{P}{\Gamma}} \, \boogy [\boogv] \, \Assign \, \translate{E}{\Gamma} \, \hoarebr{\\
  \exists \, \boogypr \, . \, \translate{P}{\Gamma}[\boogy \mapsto \boogypr] \land \boogy = {\boogypr}[\boogv \, \Assign \, \translate{E}{\Gamma} [\boogy \mapsto \boogypr]]},
  \end{multline*}
  where $\boogypr$ is some fresh variable.

  Because the case is {\tt Put}, we have just derived the Spyder triple
  $$\hoare{P} \; \sv \Put E \; \hoare{\exists \, \svpr \, . \, P[\concelem(\sv) \mapsto \svpr] \land \concelem(\sv) = E[\concelem(\sv) \mapsto \svpr]},$$
  where $\svpr$ is some free variable. 

  Let $\sigma$ be a ImpArray state such that 
  $$ \left\llbracket \exists \, \boogypr \, . \, \translate{P}{\Gamma}[\boogy \mapsto \boogypr] \land \boogy = {\boogypr}[\boogv \, \Assign \, \translate{E}{\Gamma} [\boogy \mapsto \boogypr]]\right\rrbracket_{\sigma} = t$$.
  
  Consider the Hoare term $P'$, $\exists \, \svpr \, . \, \translate{P[\concelem(\sv) \mapsto \svpr] \land \concelem(\sv) = E[\concelem(\sv) \mapsto \svpr]}{\Gamma}$, or equivalently,
  $$\exists \, \boogvpr \, . \, \translate{P}{\Gamma}[\boogy[\boogv] \mapsto \boogvpr] \land \boogy[\boogv] = \translate{E}{\Gamma}[\boogy[\boogv] \mapsto \boogvpr].$$

  We claim that $\llbracket P' \rrbracket_{\sigma} = t$. Since $P$ is well-formed with respect to $\Gamma$, and $\Gamma(\sx) = \sy$, it must be the case that the substitution of $\boogy \mapsto \boogypr$ only affects
  translations of $\concelem(\sv)$. As a result, if $\boogypr$ is an (array) witness for  
  $$ \left\llbracket \exists \, \boogypr \, . \, \translate{P}{\Gamma}[\boogy \mapsto \boogypr] \land \boogy = {\boogypr}[\boogv \, \Assign \, \translate{E}{\Gamma} [\boogy \mapsto \boogypr]]\right\rrbracket_{\sigma}, $$

  we can use the value $\boogy[\boogv]$ as a (variable) witness for $P'$.

  Since $\llbracket P' \rrbracket_{\sigma} = t$, we can apply {\tt Consequence} to obtain the triple

  $$ \hoarebr{\translate{P}{\Gamma}} \translate{\sv \, \Put \, E}{\Gamma} \, \hoarebr{P'}. $$

  \item Inductive case, in which the last step of the derivation is {\tt For:} $\hoare{ P } \; \concfor \; \pair{\sx}{\sy} B_i \; \hoare{P}$, where $P$ is of the form $\concforeach \pair{\sx}{\sy} P_i$. S

  At a high-level, this rule is introducing a quantification over the elements of $\sy$. This is sound because the body $B_i$ can only adjust the elements at the current iteration, because the loop cannot modify variables captured in $I$,
  and because the translated loop is guaranteed to execute exactly once for every element of $\sy$. 
  
  Let $\Gamma'$ be $\Gamma$ extended with the loop binding $\sx \mapsto \boogy$. Since $\Gamma$ is well-formed with respect to the loop, it must be the case that $\Gamma'$ is well-formed as well. Recall that the translated loop is $$ \boogx \, \Assign \, 0 \, \blocksep \, \concwhile \, (\boogx \, <\, \concsize(\boogy)) \, \translate{B_i}{\Gamma'} \, \blocksep \, \boogx \Assign \boogx + 1. $$ 
  
  Consider the translated $\concforeach$ predicate $I$
  $$\forall \, \boogxpr \, . \, 0 \leq \boogxpr < \concsize(\boogy) \, \implies  \, \translate{P_i}{\Gamma'}[\boogx \mapsto \boogxpr].$$

  We will use the {\tt While} rule with three helper predicates: intuitively, we keep three predicates around to 1) 
  quantify $I$ for previous iterations 2) safely weaken $I$ for the current iteration and 3) quantify $I$ for future iterations.
  Let $I_{pre}$ restrict $I$ up to the current iteration, 
  
  $$\forall \, \boogxpr \, . \, 0 \leq \boogxpr < \boogx \, \implies  \, \translate{P_i}{\Gamma'}[\boogx \mapsto \boogxpr].$$

  Let $I_{post}$ weaken $I$ using $\weakenprev{P_i}$ for future iterations:

  $$ \forall \, \boogxpr \, . \, \boogx < \boogxpr < \concsize(\boogy) \, \implies  \, \translate{\weakenprev{P_i}}{\Gamma'}[\boogx \mapsto \boogxpr].$$

  \begin{sloppypar}Finally, let $I_{curr}$ be the weakening of $I$ for the current iteration: $\translate{\weakenprev{P_i}}{\Gamma'}$.\end{sloppypar}

  We will use the {\tt While} rule with the combined predicate $I_{pre} \land I_{post} \land I_{curr}$ as the loop invariant, and in particular,
  we will show the following Hoare triple holds:

  $$ \hoarebr{I_{pre} \land I_{post} \land I_{curr} \land 0 \leq \boogx < \concsize(\boogy)} \, \translate{B_i}{\Gamma'} \, \blocksep \, \boogx \, \Assign \, \boogx + 1 \hoarebr{I_{pre} \land I_{post} \land I_{curr}}.$$
  
  From the inductive hypothesis we have the triple
  $$ \hoarebr{I_{curr} \land 0 \leq \boogx < \concsize(\boogy)} \, \translate{B_i}{\Gamma'} \, \hoarebr{\translate{P_i}{\Gamma'}}.$$

  Since $I_{pre} \land I_{post} \land I_{curr} \implies I_{curr}$, we apply {\tt Consequence} on the precondition to obtain
  $$ \hoarebr{I_{pre} \land I_{post} \land I_{curr} \land 0 \leq \boogx < \concsize(\boogy)} \, \translate{B_i}{\Gamma'} \, \hoarebr{\translate{P_i}{\Gamma'}}. $$

  From \autoref{lem:assign}, since the loop with $B_i$ is well-formed with respect to $\Gamma'$, it must be the case that $\boogx$ is not assigned within $B_i$. As well, since $B_i$ is restricted from writing to 
  free variables of $B_i$, the only way for $I_{pre}$ and $I_{post}$ to be invalidated by $\translate{B_i}{\Gamma'}$ is through a {tt Put}.
  Since Spyder does not have aliasing, each {\tt Put} within $B_i$ with $x$ as a target only writes to the current iteration (i.e. each {\tt Put} only invalidates $I_{curr}$).
  Furthermore, since $B_i$ does not have nested loops over $\sy$, $x$ is the only possible target to write to $\sy$, and so it must be the case that $I_{pre}$ and $I_{post}$ are not invalidated
  by $\translate{B_i}{\Gamma'}$.

  As a result, we can safely strengthen the postcondition of this triple with $I_{pre}$ and $I_{post}$:

  $$ \hoarebr{I_{pre} \land I_{post} \land I_{curr} \land 0 \leq \boogx < \concsize(\boogy)} \, \translate{B_i}{\Gamma'} \, \hoarebr{I_{pre} \land I_{post} \land \translate{P_i}{\Gamma'}}. $$

  Finally, consider the increment of $\boogx$ after the loop. Given the precondition $I_{pre} \land I_{post} \land \translate{P_i}{\Gamma'}$, we apply the {\tt Assign} rule to obtain

  \begin{multline*}\hoarebr{I_{pre} \land I_{post} \land \translate{P_i}{\Gamma'}} \, \boogx \, \Assign \\
   \boogx + 1 \, \hoarebr{\exists \, \boogv \, . \, (I_{pre} \land I_{post} \land \translate{P_i}{\Gamma'})[\boogx \mapsto \boogv] \land \boogx = \boogv + 1},\end{multline*}

  where $\boogv$ is a fresh variable. This postcondition is logically equivalent to $I_{pre} \land I_{post} \land I_{curr}$, and so we apply {\tt Consequence} and {\tt Sequence} to obtain

  $$ \hoarebr{I_{pre} \land I_{post} \land I_{curr} \land 0 \leq \boogx < \concsize(\boogy)} \, \translate{B_i}{\Gamma'} \, \blocksep \, \boogx \, \Assign \, \boogx + 1 \hoarebr{I_{pre} \land I_{post} \land I_{curr}}.$$

  Finally, we apply {\tt While} with the condition $0 \leq \boogx < \concsize(\boogy)$ to obtain the triple 

  \begin{multline*} \hoarebr{I_{pre} \land I_{post} \land I_{curr}} \, \concwhile \, (0 \leq \boogx < \concsize(\boogy)) \, \translate{B_i}{\Gamma'} \\
   \hoarebr{I_{pre} \land I_{post} \land I_{curr} \land \boogx = \concsize(\boogy)}.\end{multline*}

  From here, it remains to use {\tt Consequence}, and {\tt Sequence} to build a triple for the loop initialization.

\end{enumerate}
\end{proof}
\end{theorem}

\subsection{Soundness of Targeted Synthesis}

\begin{lemma}[Block Append]\label{lem:app}

  \[ \forall B, \, B', \, P, \, Q, \, R . \, \hoare{P} \, B \, \hoare{Q} \, \land \, \hoare{Q} \, B' \, \hoare{R} \quad \implies \hoare{P} \, B \concat B' \, \hoare{R}. \]

  \begin{proof}
    By structural induction over the arguments of $\concat$. 
  \end{proof}
\end{lemma}

\begin{theorem}

  \[ \forall \, \precond, \, \postcond, \, B, \, B' \, . \candVS \, ; \, \staleVS \, \vdash \hoare{\precond} \, B \, \hoare{\postcond} \, \redexstep \, B' \quad \implies \hoare{\precond} \, B' \, \hoare{\postcond}
  \]

  \begin{proof}
    Induction over the derivation of $\candVS \, ; \, \staleVS \, \vdash \hoare{\precond} \, B \, \hoare{\postcond} \, \redexstep \, B'.$ In all cases we show that $\hoare{\precond} \, B' \, \hoare{\postcond}$.
    \begin{enumerate}
      \item Base case, in which the last step is {\tt Synth-Base}: $\candVS \; ; \; \staleVS \; \vdash \; \hoare{\precond} \; \blockbase \; \hoare{\postcond} \, \redexstep \; B $ Because a side-condition for {\tt Synth-Base} is $\hoare{\precond} \, B \, \hoare{\postcond}$, this is trivially true.
      \item \begin{sloppypar}Base case, in which the last step is {\tt Synth-Loop}: $\candVS \; ; \; \staleVS \; \vdash \; \hoare{\precond} \; \blockbase \; \hoare{\concforeach \set{\pair{\svi}{\sui}} \phi \Invand \postcond} \, \redexstep \; B$. This is true from the inductive hypothesis.\end{sloppypar}
      \item Recursive case, in which the last step is {\tt Consequence}: $\candVS \; ; \; \staleVS \; \vdash \; \hoare{\precond} \; B \; \hoare{\postcond} \, \redexstep \; B' \concat  B''$. From \autoref{lem:app} and the inductive hypothesis, it is the case that $\hoare{\precond} \, B' \concat  B'' \hoare{\postcond}$.
      \item Recursive case, in which the last step is {\tt Assign}: $\candVS \; ; \; \staleVS \; \vdash \; \hoare{\precond} \; \sv \; \Assign \; E \; \blocksep B \; \hoare{\postcond} \, \redexstep \; \sv \; \Assign \; E \; \blocksep B'$. We apply the Hoare rule for {\tt Assign} to the inductive hypothesis.
      \item Recursive case, in which the last step is {\tt Put}: $\candVS \; ; \; \staleVS \; \vdash \; \hoare{\precond} \; \sv \; \Put \; E \; \blocksep B \; \hoare{\postcond} \, \redexstep \; \sv \; \Put \; E \; \blocksep B'.$ This is analogous to {\tt Assign}.
      \item Recursive case, in which the last step is one of the Extension rules. These are all trivially sound from the inductive hypothesis.
      \item Recursive case, in which the last step is {\tt Foreach-Specialize}: 
      \begin{multline*}\globalVS \; \vdash \; \hoare{\concforeach \set{\pair{\svi}{\sui}} \; \phi \Invand \postcond } \; \concfor \; \set{\pair{\sxi}{\syi}} B_i \; \blocksep B  \; \hoare{\concforeach \set{\pair{\svi}{\sui}} \; \phi \Invand \postcond} \; \redexstep \\ \concfor \set{\pair{\sxi}{\syi}} (B_{pre} \concat B_i') \; \blocksep \; B'.\end{multline*}
        In this case, we use the inductive hypothesis to establish the triple for $B_i'$. Next, we use the inductive hypothesis and \autoref{lem:app} to establish the triple for $B_i'$ and $B_{pre}$:
        $$ \hoare{\weakenprev{\postcond}} \, (B_{pre} \, \concat \, B_i') \, \hoare{\postcond}.$$
        On this, we apply the {\tt For} Hoare logic rule to introduce the $\concforeach$ term, and we appeal to the inductive hypothesis for the remainder $B'$.
      \item \begin{sloppypar}Recursive case, in which the last step is {\tt Conditional}: $\globalVS \; \vdash \; \hoare{\postcond} \; \concif \; E \; \concthen \; B_t \; \concelse \; B_f \; \blocksep B \; \hoare{\postcond} \,  \redexstep \concif \; E \; \concthen \; B_t' \; \concelse \; B_f' \; \blocksep B'$. This follows from the inductive hypothesis and the {\tt Conditional} Hoare rule.\end{sloppypar}
    \end{enumerate}
  \end{proof}
\end{theorem}

\subsection{Soundness of Targetted Synthesis}\label{app:proof-synth-sound}

\begin{theorem}

  \[ \forall \, \precond, \, \postcond, \, B, \, B' \, . \candVS \, ; \, \staleVS \, \vdash \hoare{\precond} \, B \, \hoare{\postcond} \, \redexstep \, B' \quad \implies \hoare{\precond} \, B' \, \hoare{\postcond}
  \]

  \begin{proof}
    Induction over the derivation of $\candVS \, ; \, \staleVS \, \vdash \hoare{\precond} \, B \, \hoare{\postcond} \, \redexstep \, B'.$ In all cases we show that $\hoare{\precond} \, B' \, \hoare{\postcond}$.
    \begin{enumerate}
      \item Base case, in which the last step is {\tt Synth-Base}: $\candVS \; ; \; \staleVS \; \vdash \; \hoare{\precond} \; \blockbase \; \hoare{\postcond} \, \redexstep \; B $ Because a side-condition for {\tt Synth-Base} is $\hoare{\precond} \, B \, \hoare{\postcond}$, this is trivially true.
      \item \begin{sloppypar}Base case, in which the last step is {\tt Synth-Loop}: $\candVS \; ; \; \staleVS \; \vdash \; \hoare{\precond} \; \blockbase \; \hoare{\concforeach \set{\pair{\svi}{\sui}} \phi \Invand \postcond} \, \redexstep \; B$. This is true from the inductive hypothesis.\end{sloppypar}
      \item Recursive case, in which the last step is {\tt Consequence}: $\candVS \; ; \; \staleVS \; \vdash \; \hoare{\precond} \; B \; \hoare{\postcond} \, \redexstep \; B' \concat  B''$. From \autoref{lem:app} and the inductive hypothesis, it is the case that $\hoare{\precond} \, B' \concat  B'' \hoare{\postcond}$.
      \item Recursive case, in which the last step is {\tt Assign}: $\candVS \; ; \; \staleVS \; \vdash \; \hoare{\precond} \; \sv \; \Assign \; E \; \blocksep B \; \hoare{\postcond} \, \redexstep \; \sv \; \Assign \; E \; \blocksep B'$. We apply the Hoare rule for {\tt Assign} to the inductive hypothesis.
      \item Recursive case, in which the last step is {\tt Put}: $\candVS \; ; \; \staleVS \; \vdash \; \hoare{\precond} \; \sv \; \Put \; E \; \blocksep B \; \hoare{\postcond} \, \redexstep \; \sv \; \Put \; E \; \blocksep B'.$ This is analogous to {\tt Assign}.
      \item Recursive case, in which the last step is one of the Extension rules. These are all trivially sound from the inductive hypothesis.
      \item Recursive case, in which the last step is {\tt Foreach-Specialize}: 
      \begin{multline*}\globalVS \; \vdash \; \hoare{\concforeach \set{\pair{\svi}{\sui}} \; \phi \Invand \postcond } \; \concfor \; \set{\pair{\sxi}{\syi}} B_i \; \blocksep B  \; \hoare{\concforeach \set{\pair{\svi}{\sui}} \; \phi \Invand \postcond} \; \redexstep \\ 
      \concfor \; \set{\pair{\sxi}{\syi}} (B_{pre} \concat B_i') \; \blocksep \; B'.\end{multline*}
        In this case, we use the inductive hypothesis to establish the triple for $B_i'$. Next, we use the inductive hypothesis and \autoref{lem:app} to establish the triple for $B_i'$ and $B_{pre}$:
        $$ \hoare{\weakenprev{\postcond}} \, (B_{pre} \, \concat \, B_i') \, \hoare{\postcond}.$$
        On this, we apply the {\tt For} Hoare logic rule to introduce the $\concforeach$ term, and we appeal to the inductive hypothesis for the remainder $B'$.
      \item \begin{sloppypar}Recursive case, in which the last step is {\tt Conditional}: $\globalVS \; \vdash \; \hoare{\postcond} \; \concif \; E \; \concthen \; B_t \; \concelse \; B_f \; \blocksep B \; \hoare{\postcond} \,  \redexstep \concif \; E \; \concthen \; B_t' \; \concelse \; B_f' \; \blocksep B'$. This follows from the inductive hypothesis and the {\tt Conditional} Hoare rule.\end{sloppypar}
    \end{enumerate}
  \end{proof}
\end{theorem}